\documentclass[aps,english,superscriptaddress,preprintnumbers,reprint,footinbib,amsmath,amssymb,prl]{revtex4-1}
\usepackage[version=3]{mhchem}
\usepackage{graphicx}
\usepackage{dcolumn}
\usepackage{bm}
\usepackage{hyperref}
\hypersetup{
    colorlinks=true,
    linkcolor=blue,
    filecolor=blue,
    urlcolor=blue,
   citecolor=blue,
}
\usepackage{enumitem}

\begin{document}


\title{Topological isoconductance signatures in Majorana nanowires}

\author{L. S. Ricco}
\email[corresponding author:]{luciano.ricco@unesp.br}
\affiliation{S\~ao Paulo State University (Unesp), School of Engineering, Department of Physics and Chemistry, 15385-000, Ilha Solteira-SP, Brazil}

\author{J. E. Sanches}
\affiliation{S\~ao Paulo State University (Unesp), School of Engineering, Department of Physics and Chemistry, 15385-000, Ilha Solteira-SP, Brazil} 

\author{Y. Marques}
\affiliation{ITMO University, St.~Petersburg 197101, Russia}

\author{M. de Souza}
\affiliation{S\~ao Paulo State University (Unesp), IGCE, Department of Physics, 13506-970, Rio Claro-SP, Brazil}

\author{M. S. Figueira}
\affiliation{Instituto de F\'{i}sica, Universidade Federal Fluminense, 24210-340, Niter\'{o}i, Rio de Janeiro, Brazil}

\author{I. A. Shelykh}
\affiliation{ITMO University, St.~Petersburg 197101, Russia}
\affiliation{Science Institute, University of Iceland, Dunhagi-3, IS-107,
Reykjavik, Iceland}

\author{A. C. Seridonio}
\email[corresponding author: ]{acfseridonio@gmail.com}
\affiliation{S\~ao Paulo State University (Unesp), School of Engineering, Department of Physics and Chemistry, 15385-000, Ilha Solteira-SP, Brazil}
\affiliation{S\~ao Paulo State University (Unesp), IGCE, Department of Physics, 13506-970, Rio Claro-SP, Brazil}

\date{\today}

\begin{abstract} 
We consider transport properties of a hybrid device composed by a quantum dot placed between normal and superconducting reservoirs, and coupled to a Majorana nanowire: a topological superconducting segment hosting Majorana zero-modes at the opposite ends. It is demonstrated that if topologically protected (nonoverlapping) Majorana zero-modes are formed in the system, zero-bias Andreev conductance through the dot exhibits \textit{isoconductance} profiles with the shape depending on the spin asymmetry of the coupling between a dot and a topological superconductor. Otherwise, for the topologically trivial situation corresponding to the formation of Andreev bound states, the conductance is insensitive to the spin polarization and the \textit{isoconductance} signatures disappear. This allows to propose an experimental protocol for distinguishing between isolated Majorana zero-modes and Andreev bound states.     
\end{abstract}


\maketitle

\section{Introduction}
In last years, the seek for the so-called Majorana zero-modes (MZMs) has become one of the hottest research fields in the condensed matter physics~\cite{RevMajoranaAlicea,RevMajoranaFranz,RevMajoranaAguado}. Besides fundamental interest, the unambiguous experimental detection of these exotic non-Abelian excitations is considered to be the first step towards the realization of a fault-tolerant topologically protected quantum qubit~\cite{Kitaev2001,Kitaev2003,RevNonabelian2008}. Currently, there exist a plethora of theoretical proposals of geometries where MZMs can emerge~\cite{RevMajoranaAguado}. One of the most promising alternatives is the system consisting of a segment of a quasi-one-dimensional semiconducting nanowire with strong Rashba spin-orbit (SO) coupling, brought in contact with a \textit{s-}wave superconductor and placed into external longitudinal magnetic field. 

In this setup, the proximitized nanowire is driven into the regime of unusual \textit{p-}wave superconductivity  and thereafter, if the value of the magnetic field exceeds the critical one, reaches the topological phase with MZMs appearing at the edges~\cite{LutchynPRL2010,OregPRL2010}. The experimental signature of the presence of the isolated MZMs in these so-called Majorana nanowires~\cite{Zhang2019} is the robust zero-bias peak (ZBP), appearing in tunneling spectroscopy probe measurements~\cite{MourikScience2012,KrogstrupNatMater2015,AlbrechtNature2009,DengScience2016,DengPRB2018,ZhangNatNanotech2018,ZhangNature2018,LutchynReviewMat2018,Zhang2019}. Unfortunately, other mechanisms can be responsible for the appearance of  ZBPs, as for instance the formation of zero-energy Andreev bound states (ABSs)~\cite{KellsPRB2012,LiuPRB2017,LiuPRB2018,HellPRB2018,Marra2019,LaiPRB2019,ChenPRL2019,Pan2020}. In spite of both recent theoretical and experimental efforts to distinguish between the cases of topologically protected MZMs and topologically trivial ABSs~\cite{DJClarcke2017,Prada2017,DengPRB2018,Penaranda2018,Avila2019,Ricco2019}, there is still no satisfactory solution of the problem, and the deadlock remains on the table.

\begin{figure}[t]
	\centerline{\includegraphics[width=3.0in,keepaspectratio]{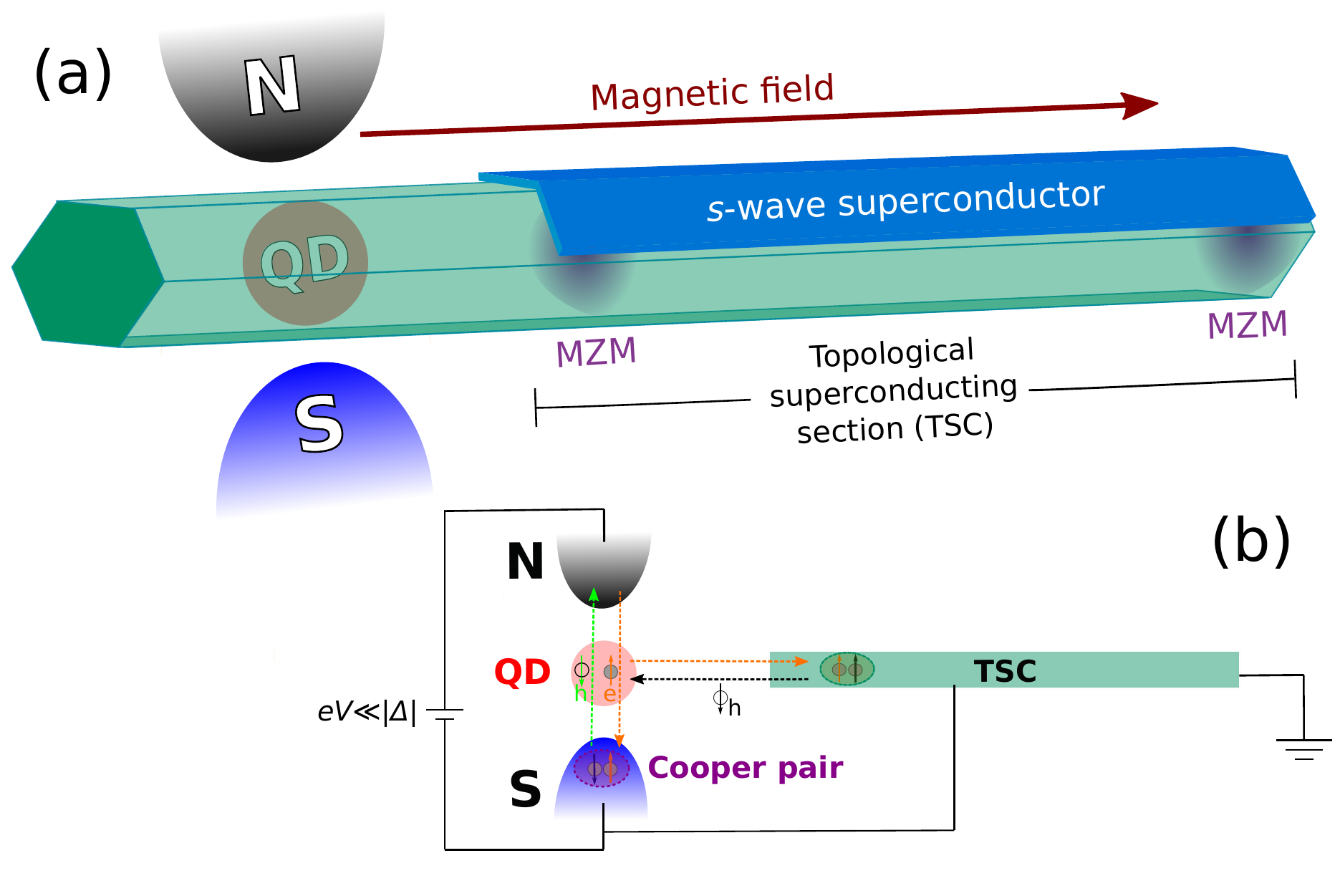}}
	\caption{\label{fig:Device}(a) Sketch of the considered setup. It consists of a QD coupled to normal (N) and superconducting (S) leads and a segment of a semiconductor nanowire covered by an \textit{s}-wave superconductor layer. In the presence of an external magnetic field parallel to the wire, the latter is driven into a topological superconducting state, with Majorana zero modes (MZMs) formed at its opposite ends. (b) The scheme illustrating spin-dependent transport channels in the system. Finite bias voltage $eV$ is applied between superconducting (S) and normal (N) reservoirs. An incoming electron from the normal reservoir with a certain spin is injected into the QD and is reflected back as a hole. In the same time, a Cooper pair is formed either in the superconducting reservoir, where it has ordinary \textit{s}-wave character, or in the TSC, where it has a \textit{p}-wave symmetry. The interplay between the transport through S and TSC terminals defines the spin orientation of the reflected hole with respect to the spin of the incoming electron.}
\end{figure}

In the current work, we theoretically propose a new protocol to differentiate between isolated MZMs corresponding to Majorana bound states (MBS), and overlapping MZMs, corresponding to ABS~\cite{LiuPRB2017,RiccoPRB2019,Pan2020} by analyzing the Andreev current through a quantum dot (QD) placed between metallic (N) and superconducting (S) reservoirs and coupled to a topological superconducting nanowire (TSC) hosting MZMs at the opposite ends (Majorana nanowire), see Fig.~\ref{fig:Device}~\cite{Baraski2016,Silva2016,Gorski2018,ZienkiewiczarXiv2019}. For the ideal situation of nonoverlapping MZMs, Andreev conductance profiles reveal strong dependence on the parameter which characterizes the spin asymmetry of the coupling between the QD and the TSC. More specifically, the zero-bias Andreev conductance as a function of both the gate-voltage defining the position of the energy level of the QD and the strength of the hybridization between the QD and superconducting lead exhibits topological \textit{isoconductance} lines. Their shape strongly depends on the spin asymmetry in the system. However, for the case of the ABS corresponding to the overlapping MZMs, the sub-gap Andreev conductance becomes spin-independent, and the aforementioned \textit{isoconductance} profiles disappear.

\section{Methods}
\subsection{Theoretical model}

To describe transport properties of the system sketched in Fig.~\ref{fig:Device}, we use the following Anderson-type Hamiltonian~\cite{Anderson,Baraski2016,Gorski2018}:
\begin{equation}
H = \sum_{\alpha=N,S} (H_{\alpha} + H_{\alpha-QD}) + H_{QD} + H_{TSC}, \label{eq:H}
\end{equation}
where $H_{N}=\sum_{\boldsymbol{k}\sigma}\varepsilon_{\boldsymbol{k}}^{N}c_{N\boldsymbol{k}\sigma}^{\dagger}c_{N\boldsymbol{k}\sigma}$ and $H_{S}=\sum_{\boldsymbol{k}\sigma}\varepsilon_{\boldsymbol{k}}^{S}c_{S\boldsymbol{k}\sigma}^{\dagger}c_{S\boldsymbol{k}\sigma}-\sum_{\boldsymbol{k}}(\Delta c_{S\boldsymbol{k}\uparrow}^{\dagger}c_{S-\boldsymbol{k}\downarrow}^{\dagger} + \text{h.c.})$ represent the N and S reservoirs, respectively, with electron energies $\varepsilon_{\boldsymbol{k}}^{\alpha}$, spin $\sigma=\uparrow,\downarrow$ and superconducting energy gap $\Delta$. $H_{\alpha-QD}=\sum_{\boldsymbol{k}\sigma}V_{\alpha \boldsymbol{k}\sigma}(c_{\alpha \boldsymbol{k} \sigma}^{\dagger}d_{\sigma} + \text{h.c.})$ stands for the hybridization between N(S) reservoir and the QD, characterized by the coupling strength $V_{\alpha \boldsymbol{k}\sigma}$. The QD is described by the Hamiltonian $H_{QD} =  \sum_{\sigma}\varepsilon_{d\sigma}d_{\sigma}^{\dagger}d_{\sigma}+Un_{d\uparrow}n_{d\downarrow}$,  corresponding to a pair of nondegenerate energy levels with the energies $\varepsilon_{d\sigma}=eV_{g}-\sigma V_{Z}$, that can be tuned by a tunnel gate $eV_{g}$ in presence of an external magnetic field inducing the Zeeman splitting $V_{Z}$, and $U$ corresponds to the Coulomb repulsion between electrons with opposite spins.

The TSC section can be modeled by the following low-energy effective spinless Hamiltonian~\cite{Hoffman2017,Prada2017}:
\begin{equation}
H_{TSC} = \imath\varepsilon_{M}\gamma_{L}\gamma_{R}+\sum_{\sigma}\left(\lambda_{\sigma}d_{\sigma}-\lambda_{\sigma}^{*}d_{\sigma}^{\dagger}\right)\gamma_{L}, \label{eq:H_TSC}
\end{equation}
where hermitian operators $\gamma_{i}=\gamma_{i}^{\dagger}$ describe the MZMs localized at the opposite ends of the TSC segment [marked in purple in Fig.~\ref{fig:Device}(a)] ~\cite{RevMajoranaFranz,RevMajoranaAguado}. The parameter $\varepsilon_{M}$ describes the overlap between the opposite MZM and thus governs the degree of the nonlocality in the system. The increase of $\varepsilon_M$ corresponds to the crossover from highly nonlocal isolated MBSs to trivial ABSs. The Hamiltonian \ref{eq:H_TSC} can be rewritten in the regular spinless fermionic basis by using the transformation $\gamma_{L}=\frac{1}{\sqrt{2}}(f + f^{\dagger})$ and $\gamma_{R}=\frac{\imath}{\sqrt{2}}(f^{\dagger}-f)$~\cite{RevMajoranaAguado,RiccoSciReports}, with $f^{\dagger}(f)$ being nonlocal fermions with ordinary Fermi-Dirac statistics.

It should be specifically stressed that although the TSC section hosting MZMs is effectively  spinless~\cite{Kitaev2001,Baranger2011,DJClarcke2017,RiccoOscillations2018}, the coupling of the MZMs to the QD depends on the spin state of the latter, and can be accounted for by introduction of the polarization parameter $p \in [0,1]$~\cite{Gorski2018,Gorski2019}, so that $\lambda_{\uparrow}=\lambda(1-p)$ and $\lambda_{\downarrow}=\lambda p$, where $\lambda$ stands for the maximal coupling amplitude. This tunable parameter depends on the effective distance between the QD and the TSC segment and the strength of the spin orbit coupling in the semiconductor nanowire, as it was shown by Hoffman \textit{et al.}~\cite{Hoffman2017}.

Since we are interested in the sub-gap Andreev transport features through the QD and its relation with the MZMs, we restrict ourselves to the limiting case of large superconducting gap $|\Delta|\rightarrow \infty$~\cite{Tanaka2007,Baraski2013,Gorski2018}. It is well known that in this regime the S lead induces static \textit{s}-wave pairing in the QD due to proximity effect. This allows to trace out the S lead from the Hamiltonian by using the substitution $H_{S} + H_{S-QD}\approx -\Gamma_{S}(d_{\uparrow}^{\dagger}d_{\downarrow}^{\dagger} + \text{h.c.})$~\cite{Oguri2004,Bauer2007,Rodero2011,MaskaPRB2017}, where $\Gamma_{S}=\pi \sum_{\boldsymbol{k}}|V_{S \boldsymbol{k} \sigma}|^{2}\delta(\omega - \varepsilon_{\boldsymbol{k}}^{\alpha})$, and in Hartree-Fock approximation~\footnote{Away from Kondo regime~\cite{Baraski2016,LeePRB,Zitko2015}, the effects of Coulomb blockade in the energy spectrum of the QD coupled to both S and N leads are well-described within the following self-consistent Hartree-Fock approximation~\cite{Bruus,LeeNanotech,Prada2017,Rodero2012}: $Un_{d\downarrow}n_{d\uparrow}\approx  U(\langle n_{d\downarrow}\rangle n_{d\uparrow}+n_{d\downarrow}\langle n_{d\uparrow}\rangle-\langle d_{\downarrow}d_{\uparrow}\rangle d_{\uparrow}^{\dagger}d_{\downarrow}^{\dagger}
-\langle d_{\uparrow}^{\dagger}d_{\downarrow}^{\dagger}\rangle d_{\downarrow}d_{\uparrow}) + \text{const.}$, where $\langle n{}_{d\sigma}\rangle=(-\frac{1}{\pi})\int_{-\infty}^{0}d\omega\text{Im}[\langle\langle d_{\sigma} d_{\sigma}^{\dagger}\rangle\rangle]$ and $\langle d_{\bar{\sigma}}^{\dagger}d_{\sigma}^{\dagger}\rangle=(-\frac{1}{\pi})\int_{-\infty}^{0}d\omega\text{Im}[\langle\langle d_{\bar{\sigma}}^{\dagger};d_{\sigma}^{\dagger}\rangle\rangle]$ are the average occupation and \textit{s}-wave pairing in the QD, respectively. Both quantities were computed with self-consistent numerical calculations.},
the system Hamiltonian given by Eq.~(\ref{eq:H}) can be rewritten as:  
\begin{eqnarray}
H & = & H_{N} + H_{N-QD} + \sum_{\sigma}\tilde{\varepsilon}_{d\sigma}d_{\sigma}^{\dagger}d_{\sigma} -(\tilde{\Gamma}_{S} d_{\uparrow}^{\dagger}d_{\downarrow}^{\dagger} + \text{h.c.})  \nonumber \\ & + & H_{TSC}, \label{eq:Heff}   
\end{eqnarray}
where $\tilde{\varepsilon}_{d\sigma}=\varepsilon_{d\sigma} + U\langle n_{d \sigma} \rangle$ and  $\tilde{\Gamma}_{S} = \Gamma_{S} + U\langle d_{\downarrow} d_{\uparrow} \rangle$.  

\subsection{Sub-gap Andreev conductance}
 At very low temperatures, when the bias-voltage $eV$ applied between the normal and superconducting reservoirs is smaller than the superconducting energy gap in the S lead ($|eV|\ll \Delta$), the electronic transport takes place exclusively due to the process of Andreev reflection ~\cite{Andreev1964}. At zero-temperature, the corresponding differential Andreev conductance can be calculated as~\cite{Gorski2018,Krawiec2003,Baraski2016}:
\begin{equation}
G_{A}(V)= \frac{2e^{2}}{h}[\mathcal{T}_{A}(\omega = -eV) + \mathcal{T}_{A}(\omega = eV)], \label{eq:G_A}    
\end{equation}
where $eV\equiv \mu_{N}-\mu_{S}$ and 
\begin{equation}
\mathcal{T}_{A}(\omega) = \Gamma_{N}^{2}\sum_{\sigma}[|\langle \langle d_{\sigma}^{\dagger}; d_{\bar{\sigma}}^{\dagger} \rangle \rangle |^{2} ]
\end{equation}
is the sub-gap transmittance due to Andreev reflection processes, which depends on the anomalous Green's functions $\langle \langle d_{\bar{\sigma}}^{\dagger}; d_{\sigma}^{\dagger} \rangle \rangle$ in the spectral domain $\omega$, with $\Gamma_{N}=\pi \sum_{\boldsymbol{k}}|V_{N \boldsymbol{k} \sigma}|^{2}\delta(\omega - \varepsilon_{\boldsymbol{k}}^{\alpha})$ being effective broadening of the QD energy levels.
%
%
\subsection{Green's functions calculation}
In order to get the anomalous Green's functions related to $\mathcal{T}_{A}$, as well as the usual Green's functions of the QD $\langle \langle d_{\sigma}; d_{\sigma}^{\dagger} \rangle \rangle$, we apply the equation-of-motion technique~\cite{Jauho,Bruus}, resulting in the following equation:  $\omega\langle\langle A_{i\sigma};B_{j\sigma'}\rangle\rangle=\langle[A_{i\sigma},B_{j\sigma'}]_{+}\rangle+\langle\langle[A_{i\sigma},H];B_{j\sigma'}\rangle\rangle,$ where $\omega=\omega+\imath\ 0^{+}$ is the spectral frequency, $A_{i\sigma}$ and $B_{j\sigma'}$ are usual fermionic operators belonging to the system Hamiltonian $H$ [Eq.~(\ref{eq:Heff})]. As we use Hartree-Fock approximation, the system Hamiltonian given by Eq.~(\ref{eq:Heff}) is bilinear, which allows to close the system of the equations for normal and anomalous Green functions, and represent it in the following form: 
\begin{widetext}
\begin{equation}
\hat{\boldsymbol{G}}_{\sigma}^{r}(\omega)=\begin{bmatrix}\langle\langle d_{\sigma};d_{\sigma}^{\dagger}\rangle\rangle\\
\langle\langle d_{\bar{\sigma}};d_{\sigma}^{\dagger}\rangle\rangle\\
\langle\langle d_{\sigma}^{\dagger};d_{\sigma}^{\dagger}\rangle\rangle\\
\langle\langle d_{\bar{\sigma}}^{\dagger};d_{\sigma}^{\dagger}\rangle\rangle\\
\langle\langle f;d_{\sigma}^{\dagger}\rangle\rangle\\
\langle\langle f^{\dagger};d_{\sigma}^{\dagger}\rangle\rangle
\end{bmatrix}=\begin{bmatrix}g_{\sigma}^{r}(\omega)^{-1} & 0 & 0 & \sigma\tilde{\Gamma}_{S}^{*} & \lambda_{\sigma}^{*}/\sqrt{2} & \lambda_{\sigma}^{*}/\sqrt{2}\\
0 & g_{\bar{\sigma}}^{r}(\omega)^{-1} & \bar{\sigma}\tilde{\Gamma}_{S}^{*} & 0 & \lambda_{\bar{\sigma}}^{*}/\sqrt{2} & \lambda_{\bar{\sigma}}^{*}/\sqrt{2}\\
0 & \bar{\sigma}\tilde{\Gamma}_{S} & \tilde{g}_{\sigma}^{r}(\omega)^{-1} & 0 & -\lambda_{\sigma}/\sqrt{2} & -\lambda_{\sigma}/\sqrt{2}\\
\sigma\tilde{\Gamma}_{S} & 0 & 0 & \tilde{g}_{\bar{\sigma}}^{r}(\omega)^{-1} & -\lambda_{\bar{\sigma}}/\sqrt{2} & -\lambda_{\bar{\sigma}}/\sqrt{2}\\
\lambda_{\sigma}/\sqrt{2} & \lambda_{\bar{\sigma}}/\sqrt{2} & -\lambda_{\sigma}^{*}/\sqrt{2} & -\lambda_{\bar{\sigma}}^{*}/\sqrt{2} & g_{\text{M}}^{r}(\omega)^{-1} & 0\\
\lambda_{\sigma}/\sqrt{2} & \lambda_{\bar{\sigma}}/\sqrt{2} & -\lambda_{\sigma}^{*}/\sqrt{2} & -\lambda_{\bar{\sigma}}^{*}/\sqrt{2} & 0 & \tilde{g}_{\text{M}}^{r}(\omega)^{-1}
\end{bmatrix}^{-1}\cdot\begin{bmatrix}1\\
0\\
0\\
0\\
0\\
0
\end{bmatrix},\label{eq:MatrixGsigma}
\end{equation}
\end{widetext}
where $g_{\sigma}^{r}(\omega)^{-1}=\omega-\tilde{\varepsilon}_{d\sigma}+\imath\Gamma_{\text{N}}$,  $\tilde{g}_{\sigma}^{r}(\omega)^{-1}=\omega+\tilde{\varepsilon}_{d\sigma}+\imath\Gamma_{\text{N}}$, $g_{\text{M}}^{r}(\omega)^{-1}=\omega-\varepsilon_{\text{M}}$ and $\tilde{g}_{\text{M}}^{r}(\omega)^{-1}=\omega+\varepsilon_{\text{M}}$~
\footnote{It is worth mentioning that Eq.~(\ref{eq:MatrixGsigma}) has similar shape of those found by Zienkiewicz \textit{et al.}~\cite{ZienkiewiczarXiv2019}. Such a matrix type also was computed by G\'{o}rski and Kucab~\cite{Gorski2019} without the S reservoir and by Ramos-Andrade \textit{et al.}~\cite{RamosAndrade2019} for a QD between N leads and side coupled to two TSC nanowires.}. 
\section{Results and Discussion}
In what follows, we use the value of $\Gamma_{N}$ as energy unit, and fix $U=2.0\Gamma_{N}$,  $V_{Z}=1.2\Gamma_{N}$ and $\lambda=2.0\Gamma_{N}$ for all considered cases.

\begin{figure}[t]
	\centerline{\includegraphics[width=3.6in,keepaspectratio]{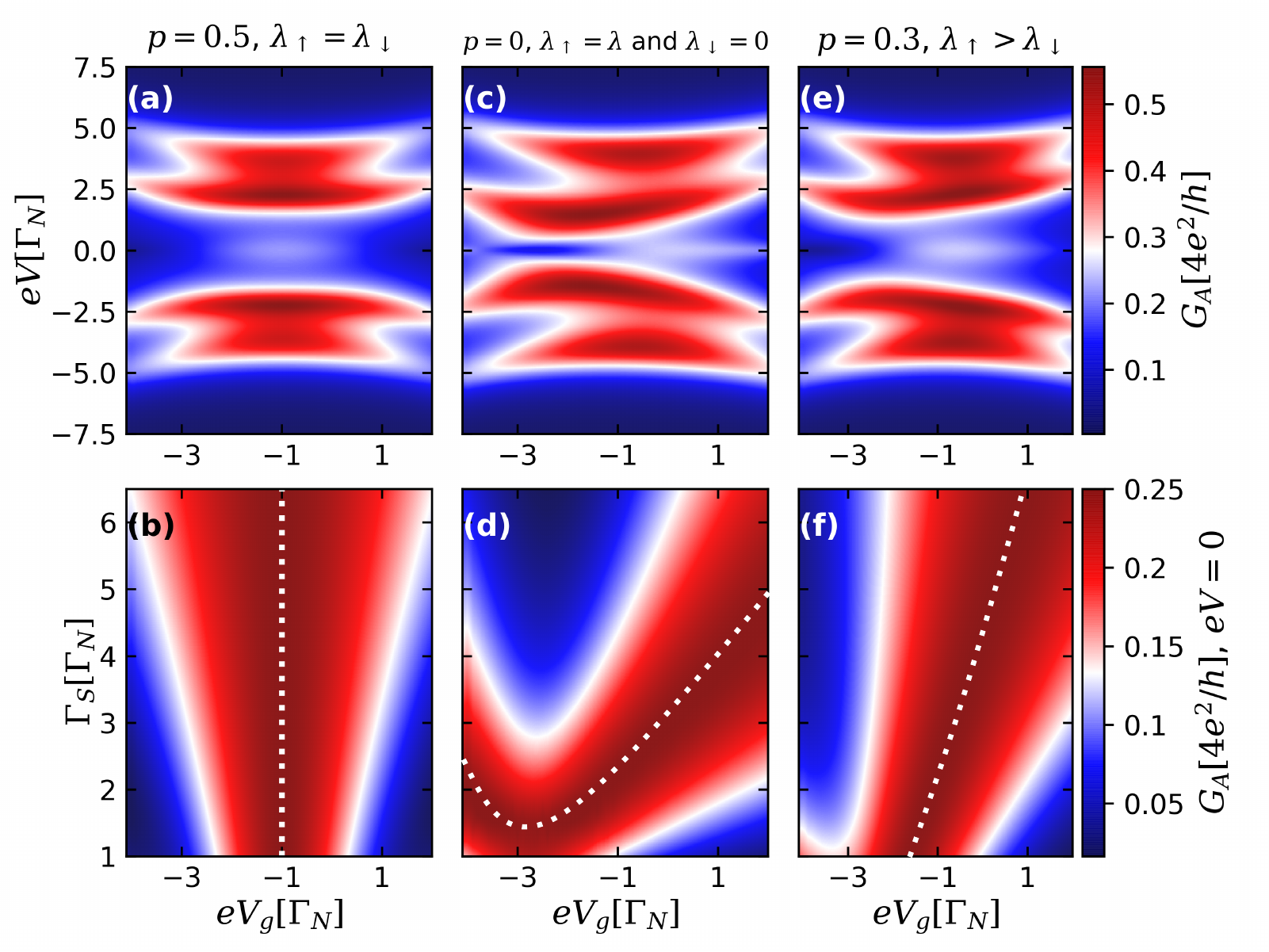}}
	\caption{Upper panels: Color scale plots of the Andreev conductance $G_A$ [Eq.~(\ref{eq:G_A})] as a function of bias voltage $eV$ and QD gate voltage $eV_{g}$, for the case of nonoverlapping MZMs, corresponding to topologically protected MBS ($\varepsilon_{\text{M}}=0$) and $\Gamma_{S}=3.0\Gamma_{N}$. Panels (a), (c) and (e)  correspond to the cases of spin-independent ($p=0.5$, $\lambda_{\uparrow}=\lambda_{\downarrow}$),  fully spin-polarized  ($p=0$, $\lambda_{\uparrow}=\lambda$ and $\lambda_{\downarrow}=0$) and intermediate ($p=0.3$, $\lambda_{\uparrow}>\lambda_{\downarrow}$) QD-TSC couplings, respectively. Lower panels: Color scale plots of Andreev conductance at zero-bias as a function of the QD-S hybridization strength $\Gamma_{S}$ and $eV_{g}$ for same  values of the parameter $p$ as in the upper panels. White dotted lines correspond to \textit{isoconductance} lines, defined by the condition that the conductance reaches its maximal value, $G_A(eV=0)=e^2/h$   
	\label{fig:Result1}}
\end{figure}

We start with the situation of nonoverlapping MZMs ($\varepsilon_{\text{M}}=0$) with a spin-independent QD-TSC coupling, putting $p=0.5$, $\lambda_{\uparrow}=\lambda_{\downarrow}=\lambda/2$. Panel (a) of Fig.~\ref{fig:Result1} shows the Andreev conductance as a function of both the bias-voltage $eV$ and the gate-voltage $eV_{g}$, shifting the position of the energy levels of the QD, for $\Gamma_{S}=3.0\Gamma_{N}$. One can clearly notice the presence of the pronounced four peak structure around $eV=0$ corresponding to the well resolved Andreev levels, appearing due to the QD-TSC coupling and splitted in the external magnetic field. Moreover, there is a visible zero-bias structure present because of the leakage of an isolated MZM into the QD \cite{VernekPRB2014,Baraski2016,ZienkiewiczarXiv2019,Baraski2013}, whose amplitude $G_{A}(eV=0)$ changes with $eV_{g}$, and reaches the maximal value of $e^{2}/h$ for $eV_{g}=-1.0\Gamma_{N}$. 

In Fig.~\ref{fig:Result1}(b) we demonstrate how Andreev conductance amplitude at zero-bias also changes as a function of both $eV_{g}$ and QD-S hybridization strength $\Gamma_{S}$ for the same case of $p=0.5$~
\footnote{Experimentally, these quantities can be continuously tuned by setting up a dual-gate geometry, with the insertion of a global back-gate in the setup, as performed by E. J. H. Lee \textit{et al.}~\cite{LeePRB}.}.
The maximal value of the conductance $e^{2}/h$ is reached along the white vertical dotted line, which we call \textit{isoconductance} line. For this particular spin-independent situation, the position of this line is defined by the condition of particle-hole symmetry, reached when $eV_{g}=-1.0\Gamma_{N}$. This condition is broken in spin asymmetric case, when $\lambda_{\uparrow}\neq\lambda_{\downarrow}$~\cite{Gorski2018}, which leads to the distortion of the \textit{isoconductance} line in the $(eV_{g},\Gamma_{S})$ space, as we shall see.    
Note also that along the \textit{isoconductance} line, the zero bias conductance does not depend on the value of $\Gamma_{S}$, so the QD becomes effectively decoupled from the S lead and the transport through it is uniquely defined by its pairing to the TSC.


\begin{figure}[t]
	\centerline{\includegraphics[width=3.6in,keepaspectratio]{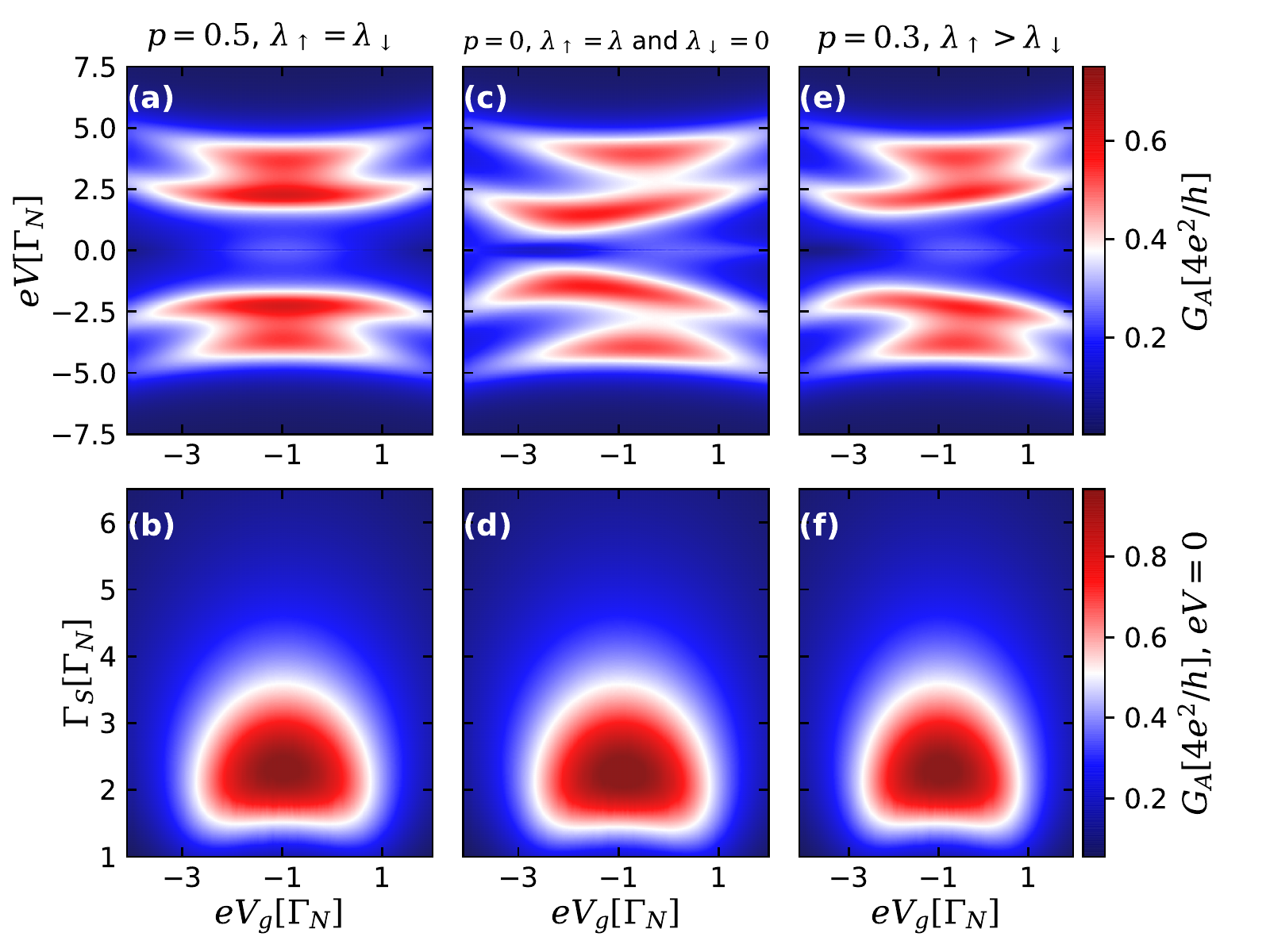}}
	\caption{Upper panels: Color scale plots of the Andreev conductance $G_A$ [Eq.~(\ref{eq:G_A})] as a function of bias voltage $eV$ and QD gate voltage $eV_{g}$, for the case of overlapping MZMs, corresponding to topologically trivial ABSs ($\varepsilon_{\text{M}}=0.05\Gamma_{N}$) and $\Gamma_{S}=3.0\Gamma_{N}$. Panels (a), (c) and (e)  correspond to the cases of spin-independent ($p=0.5$, $\lambda_{\uparrow}=\lambda_{\downarrow}$),  fully spin-polarized  ($p=0$, $\lambda_{\uparrow}=\lambda$ and $\lambda_{\downarrow}=0$) and intermediate ($p=0.3$, $\lambda_{\uparrow}>\lambda_{\downarrow}$) QD-TSC couplings, respectively. Lower panels: Color scale plots of Andreev conductance at zero-bias as a function of the QD-S hybridization strength $\Gamma_{S}$ and $eV_{g}$ for same values of parameter $p$ as in the upper panels. Note, that differently from the case of isolated MZMs illustrated by Fig.\ref{fig:Result1}, the value of the zero bias conductance $G_A(eV=0)$ can exceed $e^2/h$, and the \textit{isoconductance} lines are absent. \label{fig:Result2}}
\end{figure}

\begin{figure}[t]
	\centerline{\includegraphics[width=3.6in,keepaspectratio]{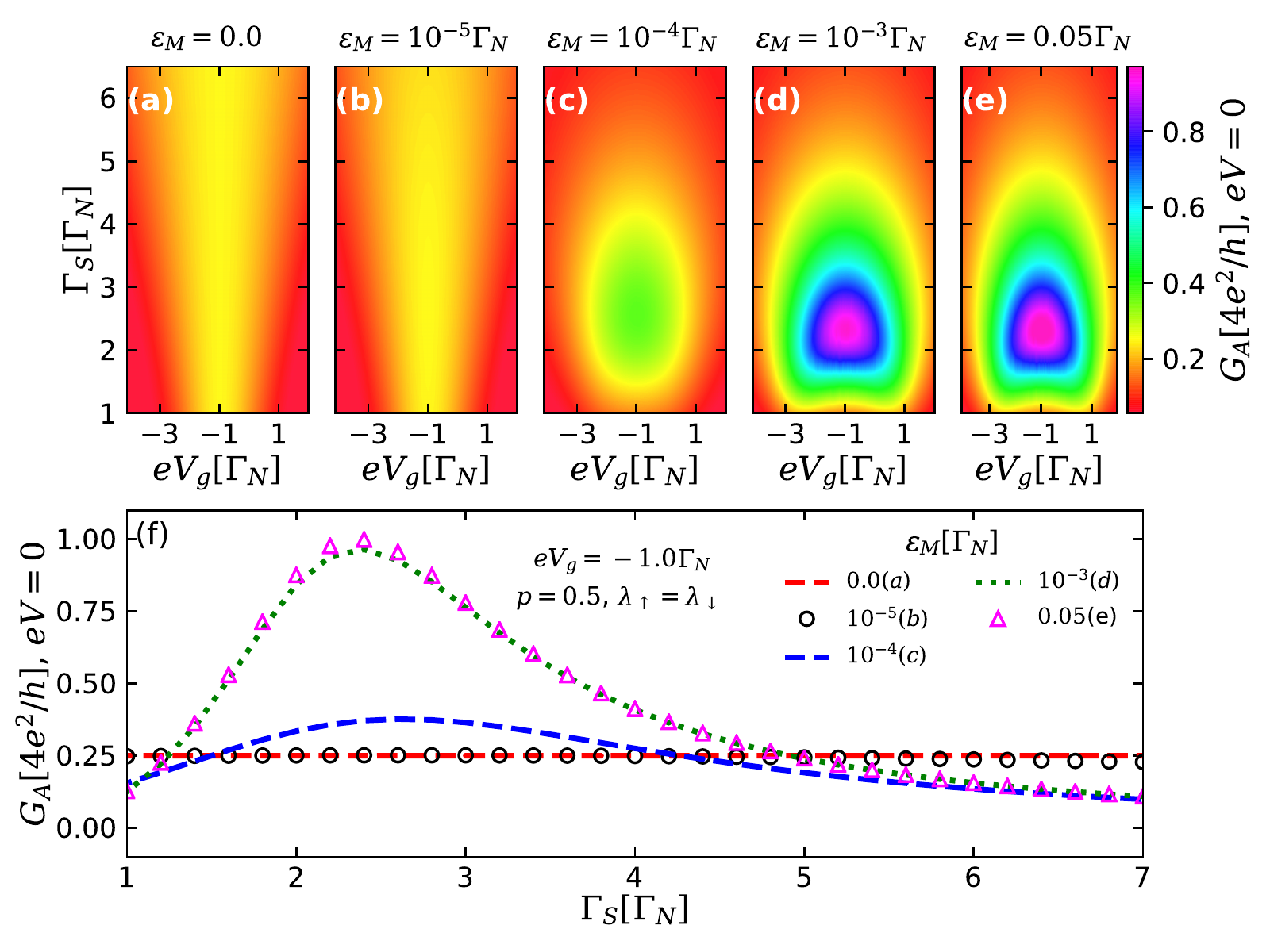}}
	\caption{(a)-(e): Color scale plots of the Andreev conductance $G_A$ [Eq.~(\ref{eq:G_A})] at zero-bias as a function of the QD-S hybridization strength $\Gamma_{S}$ and QD gate-voltage $eV_{g}$ for spin symmetric case ($p=0.5$), for five distinct values of the parameter $\varepsilon_{\text{M}}$ defining the degree of the overlap between MZMs. One clearly sees that condition $G_A(eV=0)=e^2/h$ is satisfied along the open vertical line (\textit{isoconductance} line) in the left two panels corresponding to highly isoladed MZMs, and along the closed line in the right three panels, corresponding to highly overlapping MZMs. In this latter case, the value of the conductance inside the line exceeds $e^2/h$ (f): Andreev conductance at zero-bias plotted as a function of $\Gamma_S$ with $eV_{g}=-1.0 \Gamma_{N}$, for the same values of $\varepsilon_{\text{M}}$ as in the upper panels.\label{fig:Result3}}
\end{figure}

The opposite case of fully spin polarized transport, corresponding to $p=0$, $\lambda_{\uparrow}=\lambda$ and $\lambda_{\downarrow}=0$ is illustrated by Fig.~\ref{fig:Result1}(c) and Fig.~\ref{fig:Result1}(d). The profile of the conductance as a function of the bias and gate-voltages becomes asymmetric, as it can be clearly seen in Fig.~\ref{fig:Result1}(c). Zero-bias conductance peak still appears, but the \textit{isoconductance} line defined by the condition $G_A(eV=0)=e^2/h$ is not a straight vertical line, but has a more complicated shape shown in Fig.~\ref{fig:Result1}(d).  Note that differently from the case shown in Fig.~\ref{fig:Result1}(b), the \textit{isoconductance} line has a minimum, which means that maximal value of the zero-bias conductance $e^2/h$ can not be reached below certain critical value of the coupling between the QD and the S lead. The intermediate case of $p=0.3$ is illustrated by Fig.~\ref{fig:Result1}(e) and Fig.~\ref{fig:Result1}(f). 

The comparison between the three sets of panels of Fig.~\ref{fig:Result1} allows us to conclude that the presence of an \textit{isoconductance} plateau corresponding to a vertical \textit{isoconductance} line in $eV_g,\Gamma_S$ coordinates can be considered as a hallmark of spin symmetric coupling between the QD and the TSC.

Now, let us analyze the case of overlapping MZMs ($\varepsilon_{\text{M}}=0.05\Gamma_{N}$) corresponding to the formation of topologically trivial ABS, for the cases of spin-independent ($p=0.5$), fully spin-polarized ($p=0$) and intermediary ($p=0.3$) QD-TSC couplings, as illustrated by Fig.~\ref{fig:Result2}. In the upper panels Andreev conductance as a function of the bias and gate-voltages for the fixed value of $\Gamma_S=3.0\Gamma_N$ is shown. Direct comparison with upper panels of Fig.~\ref{fig:Result1} shows, that conductance profiles are qualitatively the same for the cases of topological MBS and trivial ABS. However, if one turns to zero-bias conductance  as a function of the gate voltage $eV_g$ and QD-S lead coupling $\Gamma_S$, the results are totally different. It was already stated that for the case of the MBS (isolated MZMs, $\varepsilon_{M}=0$), the maximal value $G_A(eV=0)=e^2/h$ is reached along certain open \textit{isoconductance} lines. The situation for the case of ABS is qualitatively different. Indeed, it can be clearly seen from the lower panels of Fig.~\ref{fig:Result2} that the condition $G_A(eV=0)=e^2/h$ is reached along the closed lines, which now can not be considered as topological \textit{isoconductance} lines, as inside them the value of the conductance exceeds $e^2/h$. This remarkable difference is the signature of the formation of regular fermions and allows us to propose the experimental criterium for the distinguishing between the cases of MBS and ABS.

To study in more detail the corresponding crossover, we analyzed the zero-bias Andreev conductance as a function of $eV_g$ and $\Gamma_S$ for several values of the parameter $\varepsilon_{M}$, characterizing the overlap between the different MZMs. The results are shown in Fig.~\ref{fig:Result3}. In the panels (a)-(e) one can clearly see how an open \textit{isoconductance} line corresponding to the maximal conductance value $G_A(eV=0)=e^2/h$, observable for small $\varepsilon_M$, changes into a closed contour within which the conductance peak exceeding the value of $e^2/h$ raises. The dependence of the maximal conductance on $\Gamma_S$ for the fixed value of $eV_g$ is shown in the panel (f). Conductance plateaus, characteristic for topological MBS and corresponding to flat red solid and black open dot lines under the increase of $\varepsilon_M$ transform into non-monotonous curves corresponding to the onset of topologically trivial ABSs.

\section{Conclusions}
We have studied the sub-gap Andreev conductance $G_A$ through a quantum dot (QD) connected to metallic and superconducting leads and additionally coupled to a hybrid topological semiconducting nanowire (TSC) hosting Majorana zero-modes (MZMs) at the opposite ends. For nonoverlapping MZMs, corresponding to topological Majorana bound states (MBSs), the profiles of $G_A$ as functions of both quantum dot gate-voltage and hybridization between the dot and the superconducting reservoir reveal pronounced \textit{isoconductance} signatures, sensitive to spin selectivity of the coupling between the QD and the TSC. However, when MZMs overlap and form topologically trivial Andreev bound states,  such \textit{isoconductance} signatures disappear. This suggests that the analysis of the sub-gap Andreev conductance profiles can be employed to distinguish between the cases of authentic topologically protected MBSs and trivial ABSs.

\section*{Data Availability}
The data that support the findings of this study are available from the corresponding authors upon reasonable request.

\begin{acknowledgments}
 \section*{Acknowledgments}
 LSR acknowledges support from S\~ao Paulo Research Foundation (FAPESP), grant 2015/23539-8. ACS and MdeS acknowledge support from Brazilian National Council for Scientific and Technological Development (CNPq), grants~305668/2018-8 and 302498/2017-6, respectively. JES acknowledges support from the Coordenaç\~{a}o de Aperfeiçoamento de Pessoal de N\'{i}vel Superior - Brasil (CAPES) - Finance Code 001 (Ph.D. fellowship). MSF also acknowledges support from CNPq and CAPES funding agencies. YM and IAS acknowledge support the Ministry of Science and Higher Education of Russian Federation, goszadanie no. 2019-1246, and ITMO 5-100 Program.
\end{acknowledgments}

\section*{Author contributions}

LSR and ACS conceived the project. LSR, JES and YM carried out the calculations and plotted the figures. LSR and IAS wrote the paper with contributions from ACS, MdeS and MSF. All authors revised the manuscript.

\section*{Competing Interests}
The authors declare no competing interests.

\bibliography{apssamp}

\providecommand{\noopsort}[1]{}\providecommand{\singleletter}[1]{#1}%
\begin{thebibliography}{60}%
\makeatletter
\providecommand \@ifxundefined [1]{%
 \@ifx{#1\undefined}
}%
\providecommand \@ifnum [1]{%
 \ifnum #1\expandafter \@firstoftwo
 \else \expandafter \@secondoftwo
 \fi
}%
\providecommand \@ifx [1]{%
 \ifx #1\expandafter \@firstoftwo
 \else \expandafter \@secondoftwo
 \fi
}%
\providecommand \natexlab [1]{#1}%
\providecommand \enquote  [1]{``#1''}%
\providecommand \bibnamefont  [1]{#1}%
\providecommand \bibfnamefont [1]{#1}%
\providecommand \citenamefont [1]{#1}%
\providecommand \href@noop [0]{\@secondoftwo}%
\providecommand \href [0]{\begingroup \@sanitize@url \@href}%
\providecommand \@href[1]{\@@startlink{#1}\@@href}%
\providecommand \@@href[1]{\endgroup#1\@@endlink}%
\providecommand \@sanitize@url [0]{\catcode `\\12\catcode `\$12\catcode
  `\&12\catcode `\#12\catcode `\^12\catcode `\_12\catcode `\%12\relax}%
\providecommand \@@startlink[1]{}%
\providecommand \@@endlink[0]{}%
\providecommand \url  [0]{\begingroup\@sanitize@url \@url }%
\providecommand \@url [1]{\endgroup\@href {#1}{\urlprefix }}%
\providecommand \urlprefix  [0]{URL }%
\providecommand \Eprint [0]{\href }%
\providecommand \doibase [0]{http://dx.doi.org/}%
\providecommand \selectlanguage [0]{\@gobble}%
\providecommand \bibinfo  [0]{\@secondoftwo}%
\providecommand \bibfield  [0]{\@secondoftwo}%
\providecommand \translation [1]{[#1]}%
\providecommand \BibitemOpen [0]{}%
\providecommand \bibitemStop [0]{}%
\providecommand \bibitemNoStop [0]{.\EOS\space}%
\providecommand \EOS [0]{\spacefactor3000\relax}%
\providecommand \BibitemShut  [1]{\csname bibitem#1\endcsname}%
\let\auto@bib@innerbib\@empty
\bibitem [{\citenamefont {Alicea}(2012)}]{RevMajoranaAlicea}%
  \BibitemOpen
  \bibfield  {author} {\bibinfo {author} {\bibfnamefont {J.}~\bibnamefont
  {Alicea}},\ }\href {http://stacks.iop.org/0034-4885/75/i=7/a=076501}
  {\bibfield  {journal} {\bibinfo  {journal} {Reports on Progress in Physics}\
  }\textbf {\bibinfo {volume} {75}},\ \bibinfo {pages} {076501} (\bibinfo
  {year} {2012})}\BibitemShut {NoStop}%
\bibitem [{\citenamefont {Elliott}\ and\ \citenamefont
  {Franz}(2015)}]{RevMajoranaFranz}%
  \BibitemOpen
  \bibfield  {author} {\bibinfo {author} {\bibfnamefont {S.~R.}\ \bibnamefont
  {Elliott}}\ and\ \bibinfo {author} {\bibfnamefont {M.}~\bibnamefont
  {Franz}},\ }\href {\doibase 10.1103/RevModPhys.87.137} {\bibfield  {journal}
  {\bibinfo  {journal} {Rev. Mod. Phys.}\ }\textbf {\bibinfo {volume} {87}},\
  \bibinfo {pages} {137} (\bibinfo {year} {2015})}\BibitemShut {NoStop}%
\bibitem [{\citenamefont {Aguado}(2017)}]{RevMajoranaAguado}%
  \BibitemOpen
  \bibfield  {author} {\bibinfo {author} {\bibfnamefont {R.}~\bibnamefont
  {Aguado}},\ }\href {\doibase 10.1393/ncr/i2017-10141-9} {\bibfield  {journal}
  {\bibinfo  {journal} {Riv. Nuovo Cimento}\ }\textbf {\bibinfo {volume}
  {40}},\ \bibinfo {pages} {523} (\bibinfo {year} {2017})}\BibitemShut
  {NoStop}%
\bibitem [{\citenamefont {Kitaev}(2001)}]{Kitaev2001}%
  \BibitemOpen
  \bibfield  {author} {\bibinfo {author} {\bibfnamefont {A.~Y.}\ \bibnamefont
  {Kitaev}},\ }\href {http://stacks.iop.org/1063-7869/44/i=10S/a=S29}
  {\bibfield  {journal} {\bibinfo  {journal} {Physics-Uspekhi}\ }\textbf
  {\bibinfo {volume} {44}},\ \bibinfo {pages} {131} (\bibinfo {year}
  {2001})}\BibitemShut {NoStop}%
\bibitem [{\citenamefont {Kitaev}(2003)}]{Kitaev2003}%
  \BibitemOpen
  \bibfield  {author} {\bibinfo {author} {\bibfnamefont {A.}~\bibnamefont
  {Kitaev}},\ }\href {\doibase https://doi.org/10.1016/S0003-4916(02)00018-0}
  {\bibfield  {journal} {\bibinfo  {journal} {Annals of Physics}\ }\textbf
  {\bibinfo {volume} {303}},\ \bibinfo {pages} {2 } (\bibinfo {year}
  {2003})}\BibitemShut {NoStop}%
\bibitem [{\citenamefont {Nayak}\ \emph {et~al.}(2008)\citenamefont {Nayak},
  \citenamefont {Simon}, \citenamefont {Stern}, \citenamefont {Freedman},\ and\
  \citenamefont {Das~Sarma}}]{RevNonabelian2008}%
  \BibitemOpen
  \bibfield  {author} {\bibinfo {author} {\bibfnamefont {C.}~\bibnamefont
  {Nayak}}, \bibinfo {author} {\bibfnamefont {S.~H.}\ \bibnamefont {Simon}},
  \bibinfo {author} {\bibfnamefont {A.}~\bibnamefont {Stern}}, \bibinfo
  {author} {\bibfnamefont {M.}~\bibnamefont {Freedman}}, \ and\ \bibinfo
  {author} {\bibfnamefont {S.}~\bibnamefont {Das~Sarma}},\ }\href {\doibase
  10.1103/RevModPhys.80.1083} {\bibfield  {journal} {\bibinfo  {journal} {Rev.
  Mod. Phys.}\ }\textbf {\bibinfo {volume} {80}},\ \bibinfo {pages} {1083}
  (\bibinfo {year} {2008})}\BibitemShut {NoStop}%
\bibitem [{\citenamefont {Lutchyn}\ \emph {et~al.}(2010)\citenamefont
  {Lutchyn}, \citenamefont {Sau},\ and\ \citenamefont
  {Das~Sarma}}]{LutchynPRL2010}%
  \BibitemOpen
  \bibfield  {author} {\bibinfo {author} {\bibfnamefont {R.~M.}\ \bibnamefont
  {Lutchyn}}, \bibinfo {author} {\bibfnamefont {J.~D.}\ \bibnamefont {Sau}}, \
  and\ \bibinfo {author} {\bibfnamefont {S.}~\bibnamefont {Das~Sarma}},\ }\href
  {\doibase 10.1103/PhysRevLett.105.077001} {\bibfield  {journal} {\bibinfo
  {journal} {Phys. Rev. Lett.}\ }\textbf {\bibinfo {volume} {105}},\ \bibinfo
  {pages} {077001} (\bibinfo {year} {2010})}\BibitemShut {NoStop}%
\bibitem [{\citenamefont {Oreg}\ \emph {et~al.}(2010)\citenamefont {Oreg},
  \citenamefont {Refael},\ and\ \citenamefont {von Oppen}}]{OregPRL2010}%
  \BibitemOpen
  \bibfield  {author} {\bibinfo {author} {\bibfnamefont {Y.}~\bibnamefont
  {Oreg}}, \bibinfo {author} {\bibfnamefont {G.}~\bibnamefont {Refael}}, \ and\
  \bibinfo {author} {\bibfnamefont {F.}~\bibnamefont {von Oppen}},\ }\href
  {\doibase 10.1103/PhysRevLett.105.177002} {\bibfield  {journal} {\bibinfo
  {journal} {Phys. Rev. Lett.}\ }\textbf {\bibinfo {volume} {105}},\ \bibinfo
  {pages} {177002} (\bibinfo {year} {2010})}\BibitemShut {NoStop}%
\bibitem [{\citenamefont {Zhang}\ \emph {et~al.}(2019)\citenamefont {Zhang},
  \citenamefont {Liu}, \citenamefont {Wimmer},\ and\ \citenamefont
  {Kouwenhoven}}]{Zhang2019}%
  \BibitemOpen
  \bibfield  {author} {\bibinfo {author} {\bibfnamefont {H.}~\bibnamefont
  {Zhang}}, \bibinfo {author} {\bibfnamefont {D.~E.}\ \bibnamefont {Liu}},
  \bibinfo {author} {\bibfnamefont {M.}~\bibnamefont {Wimmer}}, \ and\ \bibinfo
  {author} {\bibfnamefont {L.~P.}\ \bibnamefont {Kouwenhoven}},\ }\href@noop {}
  {\bibfield  {journal} {\bibinfo  {journal} {arXiv e-prints}\ ,\ \bibinfo
  {eid} {arXiv:1905.07882}} (\bibinfo {year} {2019})},\ \Eprint
  {http://arxiv.org/abs/1905.07882} {arXiv:1905.07882 [cond-mat.mes-hall]}
  \BibitemShut {NoStop}%
\bibitem [{\citenamefont {Mourik}\ \emph {et~al.}(2012)\citenamefont {Mourik},
  \citenamefont {Zuo}, \citenamefont {Frolov}, \citenamefont {Plissard},
  \citenamefont {Bakkers},\ and\ \citenamefont
  {Kouwenhoven}}]{MourikScience2012}%
  \BibitemOpen
  \bibfield  {author} {\bibinfo {author} {\bibfnamefont {V.}~\bibnamefont
  {Mourik}}, \bibinfo {author} {\bibfnamefont {K.}~\bibnamefont {Zuo}},
  \bibinfo {author} {\bibfnamefont {S.~M.}\ \bibnamefont {Frolov}}, \bibinfo
  {author} {\bibfnamefont {S.~R.}\ \bibnamefont {Plissard}}, \bibinfo {author}
  {\bibfnamefont {E.~P. A.~M.}\ \bibnamefont {Bakkers}}, \ and\ \bibinfo
  {author} {\bibfnamefont {L.~P.}\ \bibnamefont {Kouwenhoven}},\ }\href
  {\doibase 10.1126/science.1222360} {\bibfield  {journal} {\bibinfo  {journal}
  {Science}\ }\textbf {\bibinfo {volume} {336}},\ \bibinfo {pages} {1003}
  (\bibinfo {year} {2012})}\BibitemShut {NoStop}%
\bibitem [{\citenamefont {Krogstrup}\ \emph {et~al.}(2015)\citenamefont
  {Krogstrup}, \citenamefont {Ziino}, \citenamefont {Chang}, \citenamefont
  {Albrecht}, \citenamefont {Madsen}, \citenamefont {Johnson}, \citenamefont
  {Nyg{\aa}rd}, \citenamefont {Marcus},\ and\ \citenamefont
  {Jespersen}}]{KrogstrupNatMater2015}%
  \BibitemOpen
  \bibfield  {author} {\bibinfo {author} {\bibfnamefont {P.}~\bibnamefont
  {Krogstrup}}, \bibinfo {author} {\bibfnamefont {N.~L.~B.}\ \bibnamefont
  {Ziino}}, \bibinfo {author} {\bibfnamefont {W.}~\bibnamefont {Chang}},
  \bibinfo {author} {\bibfnamefont {S.~M.}\ \bibnamefont {Albrecht}}, \bibinfo
  {author} {\bibfnamefont {M.~H.}\ \bibnamefont {Madsen}}, \bibinfo {author}
  {\bibfnamefont {E.}~\bibnamefont {Johnson}}, \bibinfo {author} {\bibfnamefont
  {J.}~\bibnamefont {Nyg{\aa}rd}}, \bibinfo {author} {\bibfnamefont {C.~M.}\
  \bibnamefont {Marcus}}, \ and\ \bibinfo {author} {\bibfnamefont {T.~S.}\
  \bibnamefont {Jespersen}},\ }\href {\doibase 10.1038/nmat4176} {\bibfield
  {journal} {\bibinfo  {journal} {Nat.Mater.}\ }\textbf {\bibinfo {volume}
  {14}},\ \bibinfo {pages} {1476} (\bibinfo {year} {2015})}\BibitemShut
  {NoStop}%
\bibitem [{\citenamefont {Albrecht}\ \emph {et~al.}(2016)\citenamefont
  {Albrecht}, \citenamefont {Higginbotham}, \citenamefont {Madsen},
  \citenamefont {Kuemmeth}, \citenamefont {Jespersen}, \citenamefont
  {Nyg{\aa}rd}, \citenamefont {Krogstrup},\ and\ \citenamefont
  {Marcus}}]{AlbrechtNature2009}%
  \BibitemOpen
  \bibfield  {author} {\bibinfo {author} {\bibfnamefont {S.~M.}\ \bibnamefont
  {Albrecht}}, \bibinfo {author} {\bibfnamefont {A.}~\bibnamefont
  {Higginbotham}}, \bibinfo {author} {\bibfnamefont {M.}~\bibnamefont
  {Madsen}}, \bibinfo {author} {\bibfnamefont {F.}~\bibnamefont {Kuemmeth}},
  \bibinfo {author} {\bibfnamefont {T.~S.}\ \bibnamefont {Jespersen}}, \bibinfo
  {author} {\bibfnamefont {J.}~\bibnamefont {Nyg{\aa}rd}}, \bibinfo {author}
  {\bibfnamefont {P.}~\bibnamefont {Krogstrup}}, \ and\ \bibinfo {author}
  {\bibfnamefont {C.}~\bibnamefont {Marcus}},\ }\href {\doibase
  10.1038/nature17162} {\bibfield  {journal} {\bibinfo  {journal} {Nature}\
  }\textbf {\bibinfo {volume} {531}},\ \bibinfo {pages} {206} (\bibinfo {year}
  {2016})}\BibitemShut {NoStop}%
\bibitem [{\citenamefont {Deng}\ \emph {et~al.}(2016)\citenamefont {Deng},
  \citenamefont {Vaitiekenas}, \citenamefont {Hansen}, \citenamefont {Danon},
  \citenamefont {Leijnse}, \citenamefont {Flensberg}, \citenamefont {Nyg{\r
  a}rd}, \citenamefont {Krogstrup},\ and\ \citenamefont
  {Marcus}}]{DengScience2016}%
  \BibitemOpen
  \bibfield  {author} {\bibinfo {author} {\bibfnamefont {M.~T.}\ \bibnamefont
  {Deng}}, \bibinfo {author} {\bibfnamefont {S.}~\bibnamefont {Vaitiekenas}},
  \bibinfo {author} {\bibfnamefont {E.~B.}\ \bibnamefont {Hansen}}, \bibinfo
  {author} {\bibfnamefont {J.}~\bibnamefont {Danon}}, \bibinfo {author}
  {\bibfnamefont {M.}~\bibnamefont {Leijnse}}, \bibinfo {author} {\bibfnamefont
  {K.}~\bibnamefont {Flensberg}}, \bibinfo {author} {\bibfnamefont
  {J.}~\bibnamefont {Nyg{\r a}rd}}, \bibinfo {author} {\bibfnamefont
  {P.}~\bibnamefont {Krogstrup}}, \ and\ \bibinfo {author} {\bibfnamefont
  {C.~M.}\ \bibnamefont {Marcus}},\ }\href {\doibase 10.1126/science.aaf3961}
  {\bibfield  {journal} {\bibinfo  {journal} {Science}\ }\textbf {\bibinfo
  {volume} {354}},\ \bibinfo {pages} {1557} (\bibinfo {year}
  {2016})}\BibitemShut {NoStop}%
\bibitem [{\citenamefont {Deng}\ \emph {et~al.}(2018)\citenamefont {Deng},
  \citenamefont {Vaitiek\ifmmode~\dot{e}\else \.{e}\fi{}nas}, \citenamefont
  {Prada}, \citenamefont {San-Jose}, \citenamefont {Nyg\aa{}rd}, \citenamefont
  {Krogstrup}, \citenamefont {Aguado},\ and\ \citenamefont
  {Marcus}}]{DengPRB2018}%
  \BibitemOpen
  \bibfield  {author} {\bibinfo {author} {\bibfnamefont {M.-T.}\ \bibnamefont
  {Deng}}, \bibinfo {author} {\bibfnamefont {S.}~\bibnamefont
  {Vaitiek\ifmmode~\dot{e}\else \.{e}\fi{}nas}}, \bibinfo {author}
  {\bibfnamefont {E.}~\bibnamefont {Prada}}, \bibinfo {author} {\bibfnamefont
  {P.}~\bibnamefont {San-Jose}}, \bibinfo {author} {\bibfnamefont
  {J.}~\bibnamefont {Nyg\aa{}rd}}, \bibinfo {author} {\bibfnamefont
  {P.}~\bibnamefont {Krogstrup}}, \bibinfo {author} {\bibfnamefont
  {R.}~\bibnamefont {Aguado}}, \ and\ \bibinfo {author} {\bibfnamefont {C.~M.}\
  \bibnamefont {Marcus}},\ }\href {\doibase 10.1103/PhysRevB.98.085125}
  {\bibfield  {journal} {\bibinfo  {journal} {Phys. Rev. B}\ }\textbf {\bibinfo
  {volume} {98}},\ \bibinfo {pages} {085125} (\bibinfo {year}
  {2018})}\BibitemShut {NoStop}%
\bibitem [{\citenamefont {{Zhang}}\ \emph {et~al.}(2018)\citenamefont
  {{Zhang}}, \citenamefont {{G{\"u}l}}, \citenamefont {{Conesa-Boj}},
  \citenamefont {{Zuo}}, \citenamefont {{Mourik}}, \citenamefont {{de Vries}},
  \citenamefont {{van Veen}}, \citenamefont {{van Woerkom}}, \citenamefont
  {{Nowak}}, \citenamefont {{Wimmer}}, \citenamefont {{Car}}, \citenamefont
  {{Plissard}}, \citenamefont {{Bakkers}}, \citenamefont
  {{Quintero-P{\'e}rez}}, \citenamefont {{Goswami}}, \citenamefont
  {{Watanabe}}, \citenamefont {{Taniguchi}},\ and\ \citenamefont
  {{Kouwenhoven}}}]{ZhangNatNanotech2018}%
  \BibitemOpen
  \bibfield  {author} {\bibinfo {author} {\bibfnamefont {H.}~\bibnamefont
  {{Zhang}}}, \bibinfo {author} {\bibfnamefont {{\"O}.}~\bibnamefont
  {{G{\"u}l}}}, \bibinfo {author} {\bibfnamefont {S.}~\bibnamefont
  {{Conesa-Boj}}}, \bibinfo {author} {\bibfnamefont {K.}~\bibnamefont {{Zuo}}},
  \bibinfo {author} {\bibfnamefont {V.}~\bibnamefont {{Mourik}}}, \bibinfo
  {author} {\bibfnamefont {F.~K.}\ \bibnamefont {{de Vries}}}, \bibinfo
  {author} {\bibfnamefont {J.}~\bibnamefont {{van Veen}}}, \bibinfo {author}
  {\bibfnamefont {D.~J.}\ \bibnamefont {{van Woerkom}}}, \bibinfo {author}
  {\bibfnamefont {M.~P.}\ \bibnamefont {{Nowak}}}, \bibinfo {author}
  {\bibfnamefont {M.}~\bibnamefont {{Wimmer}}}, \bibinfo {author}
  {\bibfnamefont {D.}~\bibnamefont {{Car}}}, \bibinfo {author} {\bibfnamefont
  {S.}~\bibnamefont {{Plissard}}}, \bibinfo {author} {\bibfnamefont
  {E.~P.~A.~M.}\ \bibnamefont {{Bakkers}}}, \bibinfo {author} {\bibfnamefont
  {M.}~\bibnamefont {{Quintero-P{\'e}rez}}}, \bibinfo {author} {\bibfnamefont
  {S.}~\bibnamefont {{Goswami}}}, \bibinfo {author} {\bibfnamefont
  {K.}~\bibnamefont {{Watanabe}}}, \bibinfo {author} {\bibfnamefont
  {T.}~\bibnamefont {{Taniguchi}}}, \ and\ \bibinfo {author} {\bibfnamefont
  {L.~P.}\ \bibnamefont {{Kouwenhoven}}},\ }\href {\doibase
  10.1038/s41565-017-0032-8} {\bibfield  {journal} {\bibinfo  {journal} {Nature
  Nanotechnology}\ }\textbf {\bibinfo {volume} {13}},\ \bibinfo {pages} {1748}
  (\bibinfo {year} {2018})}\BibitemShut {NoStop}%
\bibitem [{\citenamefont {Zhang}\ \emph {et~al.}(2018)\citenamefont {Zhang},
  \citenamefont {Liu}, \citenamefont {Gazibegovic}, \citenamefont {Xu},
  \citenamefont {Logan}, \citenamefont {Wang}, \citenamefont {van Loo},
  \citenamefont {Bommer}, \citenamefont {de~Moor}, \citenamefont {Car},
  \citenamefont {Op~het Veld}, \citenamefont {van Veldhoven}, \citenamefont
  {Koelling}, \citenamefont {Verheijen}, \citenamefont {Pendharkar},
  \citenamefont {Pennachio}, \citenamefont {Shojaei}, \citenamefont {Lee},
  \citenamefont {Palmstr\o{}m}, \citenamefont {Bakkers}, \citenamefont
  {Sarma},\ and\ \citenamefont {Kouwenhoven}}]{ZhangNature2018}%
  \BibitemOpen
  \bibfield  {author} {\bibinfo {author} {\bibfnamefont {H.}~\bibnamefont
  {Zhang}}, \bibinfo {author} {\bibfnamefont {C.-X.}\ \bibnamefont {Liu}},
  \bibinfo {author} {\bibfnamefont {S.}~\bibnamefont {Gazibegovic}}, \bibinfo
  {author} {\bibfnamefont {D.}~\bibnamefont {Xu}}, \bibinfo {author}
  {\bibfnamefont {J.~A.}\ \bibnamefont {Logan}}, \bibinfo {author}
  {\bibfnamefont {G.}~\bibnamefont {Wang}}, \bibinfo {author} {\bibfnamefont
  {N.}~\bibnamefont {van Loo}}, \bibinfo {author} {\bibfnamefont {J.~D.~S.}\
  \bibnamefont {Bommer}}, \bibinfo {author} {\bibfnamefont {M.~W.~A.}\
  \bibnamefont {de~Moor}}, \bibinfo {author} {\bibfnamefont {D.}~\bibnamefont
  {Car}}, \bibinfo {author} {\bibfnamefont {R.~L.~M.}\ \bibnamefont {Op~het
  Veld}}, \bibinfo {author} {\bibfnamefont {P.~J.}\ \bibnamefont {van
  Veldhoven}}, \bibinfo {author} {\bibfnamefont {S.}~\bibnamefont {Koelling}},
  \bibinfo {author} {\bibfnamefont {M.~A.}\ \bibnamefont {Verheijen}}, \bibinfo
  {author} {\bibfnamefont {M.}~\bibnamefont {Pendharkar}}, \bibinfo {author}
  {\bibfnamefont {D.~J.}\ \bibnamefont {Pennachio}}, \bibinfo {author}
  {\bibfnamefont {B.}~\bibnamefont {Shojaei}}, \bibinfo {author} {\bibfnamefont
  {J.~S.}\ \bibnamefont {Lee}}, \bibinfo {author} {\bibfnamefont {C.~J.}\
  \bibnamefont {Palmstr\o{}m}}, \bibinfo {author} {\bibfnamefont {E.~P. A.~M.}\
  \bibnamefont {Bakkers}}, \bibinfo {author} {\bibfnamefont {S.~D.}\
  \bibnamefont {Sarma}}, \ and\ \bibinfo {author} {\bibfnamefont {L.~P.}\
  \bibnamefont {Kouwenhoven}},\ }\href {\doibase 10.1038/nature26142}
  {\bibfield  {journal} {\bibinfo  {journal} {Nature}\ }\textbf {\bibinfo
  {volume} {556}},\ \bibinfo {pages} {74} (\bibinfo {year} {2018})}\BibitemShut
  {NoStop}%
\bibitem [{\citenamefont {Lutchyn}\ \emph {et~al.}(2018)\citenamefont
  {Lutchyn}, \citenamefont {Bakkers}, \citenamefont {Kouwenhoven},
  \citenamefont {Krogstrup}, \citenamefont {Marcus},\ and\ \citenamefont
  {Oreg}}]{LutchynReviewMat2018}%
  \BibitemOpen
  \bibfield  {author} {\bibinfo {author} {\bibfnamefont {R.~M.}\ \bibnamefont
  {Lutchyn}}, \bibinfo {author} {\bibfnamefont {E.~P. A.~M.}\ \bibnamefont
  {Bakkers}}, \bibinfo {author} {\bibfnamefont {L.~P.}\ \bibnamefont
  {Kouwenhoven}}, \bibinfo {author} {\bibfnamefont {P.}~\bibnamefont
  {Krogstrup}}, \bibinfo {author} {\bibfnamefont {C.~M.}\ \bibnamefont
  {Marcus}}, \ and\ \bibinfo {author} {\bibfnamefont {Y.}~\bibnamefont
  {Oreg}},\ }\href {\doibase 10.1038/s41578-018-0003-1} {\bibfield  {journal}
  {\bibinfo  {journal} {Nature Reviews Materials}\ }\textbf {\bibinfo {volume}
  {3}},\ \bibinfo {pages} {52} (\bibinfo {year} {2018})}\BibitemShut {NoStop}%
\bibitem [{\citenamefont {Kells}\ \emph {et~al.}(2012)\citenamefont {Kells},
  \citenamefont {Meidan},\ and\ \citenamefont {Brouwer}}]{KellsPRB2012}%
  \BibitemOpen
  \bibfield  {author} {\bibinfo {author} {\bibfnamefont {G.}~\bibnamefont
  {Kells}}, \bibinfo {author} {\bibfnamefont {D.}~\bibnamefont {Meidan}}, \
  and\ \bibinfo {author} {\bibfnamefont {P.~W.}\ \bibnamefont {Brouwer}},\
  }\href {\doibase 10.1103/PhysRevB.86.100503} {\bibfield  {journal} {\bibinfo
  {journal} {Phys. Rev. B}\ }\textbf {\bibinfo {volume} {86}},\ \bibinfo
  {pages} {100503} (\bibinfo {year} {2012})}\BibitemShut {NoStop}%
\bibitem [{\citenamefont {Liu}\ \emph {et~al.}(2017)\citenamefont {Liu},
  \citenamefont {Sau}, \citenamefont {Stanescu},\ and\ \citenamefont
  {Das~Sarma}}]{LiuPRB2017}%
  \BibitemOpen
  \bibfield  {author} {\bibinfo {author} {\bibfnamefont {C.-X.}\ \bibnamefont
  {Liu}}, \bibinfo {author} {\bibfnamefont {J.~D.}\ \bibnamefont {Sau}},
  \bibinfo {author} {\bibfnamefont {T.~D.}\ \bibnamefont {Stanescu}}, \ and\
  \bibinfo {author} {\bibfnamefont {S.}~\bibnamefont {Das~Sarma}},\ }\href
  {\doibase 10.1103/PhysRevB.96.075161} {\bibfield  {journal} {\bibinfo
  {journal} {Phys. Rev. B}\ }\textbf {\bibinfo {volume} {96}},\ \bibinfo
  {pages} {075161} (\bibinfo {year} {2017})}\BibitemShut {NoStop}%
\bibitem [{\citenamefont {Liu}\ \emph {et~al.}(2018)\citenamefont {Liu},
  \citenamefont {Sau},\ and\ \citenamefont {Das~Sarma}}]{LiuPRB2018}%
  \BibitemOpen
  \bibfield  {author} {\bibinfo {author} {\bibfnamefont {C.-X.}\ \bibnamefont
  {Liu}}, \bibinfo {author} {\bibfnamefont {J.~D.}\ \bibnamefont {Sau}}, \ and\
  \bibinfo {author} {\bibfnamefont {S.}~\bibnamefont {Das~Sarma}},\ }\href
  {\doibase 10.1103/PhysRevB.97.214502} {\bibfield  {journal} {\bibinfo
  {journal} {Phys. Rev. B}\ }\textbf {\bibinfo {volume} {97}},\ \bibinfo
  {pages} {214502} (\bibinfo {year} {2018})}\BibitemShut {NoStop}%
\bibitem [{\citenamefont {Hell}\ \emph {et~al.}(2018)\citenamefont {Hell},
  \citenamefont {Flensberg},\ and\ \citenamefont {Leijnse}}]{HellPRB2018}%
  \BibitemOpen
  \bibfield  {author} {\bibinfo {author} {\bibfnamefont {M.}~\bibnamefont
  {Hell}}, \bibinfo {author} {\bibfnamefont {K.}~\bibnamefont {Flensberg}}, \
  and\ \bibinfo {author} {\bibfnamefont {M.}~\bibnamefont {Leijnse}},\ }\href
  {\doibase 10.1103/PhysRevB.97.161401} {\bibfield  {journal} {\bibinfo
  {journal} {Phys. Rev. B}\ }\textbf {\bibinfo {volume} {97}},\ \bibinfo
  {pages} {161401} (\bibinfo {year} {2018})}\BibitemShut {NoStop}%
\bibitem [{\citenamefont {Marra}\ and\ \citenamefont
  {Nitta}(2019)}]{Marra2019}%
  \BibitemOpen
  \bibfield  {author} {\bibinfo {author} {\bibfnamefont {P.}~\bibnamefont
  {Marra}}\ and\ \bibinfo {author} {\bibfnamefont {M.}~\bibnamefont {Nitta}},\
  }\href@noop {} {\bibfield  {journal} {\bibinfo  {journal} {arXiv e-prints}\
  ,\ \bibinfo {eid} {arXiv:1907.05416}} (\bibinfo {year} {2019})},\ \Eprint
  {http://arxiv.org/abs/1907.05416} {arXiv:1907.05416 [cond-mat.supr-con]}
  \BibitemShut {NoStop}%
\bibitem [{\citenamefont {Lai}\ \emph {et~al.}(2019)\citenamefont {Lai},
  \citenamefont {Sau},\ and\ \citenamefont {Das~Sarma}}]{LaiPRB2019}%
  \BibitemOpen
  \bibfield  {author} {\bibinfo {author} {\bibfnamefont {Y.-H.}\ \bibnamefont
  {Lai}}, \bibinfo {author} {\bibfnamefont {J.~D.}\ \bibnamefont {Sau}}, \ and\
  \bibinfo {author} {\bibfnamefont {S.}~\bibnamefont {Das~Sarma}},\ }\href
  {\doibase 10.1103/PhysRevB.100.045302} {\bibfield  {journal} {\bibinfo
  {journal} {Phys. Rev. B}\ }\textbf {\bibinfo {volume} {100}},\ \bibinfo
  {pages} {045302} (\bibinfo {year} {2019})}\BibitemShut {NoStop}%
\bibitem [{\citenamefont {Chen}\ \emph {et~al.}(2019)\citenamefont {Chen},
  \citenamefont {Woods}, \citenamefont {Yu}, \citenamefont {Hocevar},
  \citenamefont {Car}, \citenamefont {Plissard}, \citenamefont {Bakkers},
  \citenamefont {Stanescu},\ and\ \citenamefont {Frolov}}]{ChenPRL2019}%
  \BibitemOpen
  \bibfield  {author} {\bibinfo {author} {\bibfnamefont {J.}~\bibnamefont
  {Chen}}, \bibinfo {author} {\bibfnamefont {B.~D.}\ \bibnamefont {Woods}},
  \bibinfo {author} {\bibfnamefont {P.}~\bibnamefont {Yu}}, \bibinfo {author}
  {\bibfnamefont {M.}~\bibnamefont {Hocevar}}, \bibinfo {author} {\bibfnamefont
  {D.}~\bibnamefont {Car}}, \bibinfo {author} {\bibfnamefont {S.~R.}\
  \bibnamefont {Plissard}}, \bibinfo {author} {\bibfnamefont {E.~P. A.~M.}\
  \bibnamefont {Bakkers}}, \bibinfo {author} {\bibfnamefont {T.~D.}\
  \bibnamefont {Stanescu}}, \ and\ \bibinfo {author} {\bibfnamefont {S.~M.}\
  \bibnamefont {Frolov}},\ }\href {\doibase 10.1103/PhysRevLett.123.107703}
  {\bibfield  {journal} {\bibinfo  {journal} {Phys. Rev. Lett.}\ }\textbf
  {\bibinfo {volume} {123}},\ \bibinfo {pages} {107703} (\bibinfo {year}
  {2019})}\BibitemShut {NoStop}%
\bibitem [{\citenamefont {Pan}\ and\ \citenamefont
  {Das~Sarma}(2020)}]{Pan2020}%
  \BibitemOpen
  \bibfield  {author} {\bibinfo {author} {\bibfnamefont {H.}~\bibnamefont
  {Pan}}\ and\ \bibinfo {author} {\bibfnamefont {S.}~\bibnamefont
  {Das~Sarma}},\ }\href {\doibase 10.1103/PhysRevResearch.2.013377} {\bibfield
  {journal} {\bibinfo  {journal} {Phys. Rev. Research}\ }\textbf {\bibinfo
  {volume} {2}},\ \bibinfo {pages} {013377} (\bibinfo {year}
  {2020})}\BibitemShut {NoStop}%
\bibitem [{\citenamefont {Clarke}(2017)}]{DJClarcke2017}%
  \BibitemOpen
  \bibfield  {author} {\bibinfo {author} {\bibfnamefont {D.~J.}\ \bibnamefont
  {Clarke}},\ }\href {\doibase 10.1103/PhysRevB.96.201109} {\bibfield
  {journal} {\bibinfo  {journal} {Phys. Rev. B}\ }\textbf {\bibinfo {volume}
  {96}},\ \bibinfo {pages} {201109} (\bibinfo {year} {2017})}\BibitemShut
  {NoStop}%
\bibitem [{\citenamefont {Prada}\ \emph {et~al.}(2017)\citenamefont {Prada},
  \citenamefont {Aguado},\ and\ \citenamefont {San-Jose}}]{Prada2017}%
  \BibitemOpen
  \bibfield  {author} {\bibinfo {author} {\bibfnamefont {E.}~\bibnamefont
  {Prada}}, \bibinfo {author} {\bibfnamefont {R.}~\bibnamefont {Aguado}}, \
  and\ \bibinfo {author} {\bibfnamefont {P.}~\bibnamefont {San-Jose}},\ }\href
  {\doibase 10.1103/PhysRevB.96.085418} {\bibfield  {journal} {\bibinfo
  {journal} {Phys. Rev. B}\ }\textbf {\bibinfo {volume} {96}},\ \bibinfo
  {pages} {085418} (\bibinfo {year} {2017})}\BibitemShut {NoStop}%
\bibitem [{\citenamefont {Pe\~naranda}\ \emph {et~al.}(2018)\citenamefont
  {Pe\~naranda}, \citenamefont {Aguado}, \citenamefont {San-Jose},\ and\
  \citenamefont {Prada}}]{Penaranda2018}%
  \BibitemOpen
  \bibfield  {author} {\bibinfo {author} {\bibfnamefont {F.}~\bibnamefont
  {Pe\~naranda}}, \bibinfo {author} {\bibfnamefont {R.}~\bibnamefont {Aguado}},
  \bibinfo {author} {\bibfnamefont {P.}~\bibnamefont {San-Jose}}, \ and\
  \bibinfo {author} {\bibfnamefont {E.}~\bibnamefont {Prada}},\ }\href
  {\doibase 10.1103/PhysRevB.98.235406} {\bibfield  {journal} {\bibinfo
  {journal} {Phys. Rev. B}\ }\textbf {\bibinfo {volume} {98}},\ \bibinfo
  {pages} {235406} (\bibinfo {year} {2018})}\BibitemShut {NoStop}%
\bibitem [{\citenamefont {Avila}\ \emph {et~al.}(2019)\citenamefont {Avila},
  \citenamefont {Peñaranda}, \citenamefont {Prada}, \citenamefont {San-Jose},\
  and\ \citenamefont {Aguado}}]{Avila2019}%
  \BibitemOpen
  \bibfield  {author} {\bibinfo {author} {\bibfnamefont {J.}~\bibnamefont
  {Avila}}, \bibinfo {author} {\bibfnamefont {F.}~\bibnamefont {Peñaranda}},
  \bibinfo {author} {\bibfnamefont {E.}~\bibnamefont {Prada}}, \bibinfo
  {author} {\bibfnamefont {P.}~\bibnamefont {San-Jose}}, \ and\ \bibinfo
  {author} {\bibfnamefont {R.}~\bibnamefont {Aguado}},\ }\href {\doibase
  10.1038/s42005-019-0231-8} {\bibfield  {journal} {\bibinfo  {journal}
  {Commun. Phys.}\ }\textbf {\bibinfo {volume} {2}},\ \bibinfo {pages} {133}
  (\bibinfo {year} {2019})}\BibitemShut {NoStop}%
\bibitem [{\citenamefont {Ricco}\ \emph
  {et~al.}(2019{\natexlab{a}})\citenamefont {Ricco}, \citenamefont {de~Souza},
  \citenamefont {Figueira}, \citenamefont {Shelykh},\ and\ \citenamefont
  {Seridonio}}]{Ricco2019}%
  \BibitemOpen
  \bibfield  {author} {\bibinfo {author} {\bibfnamefont {L.~S.}\ \bibnamefont
  {Ricco}}, \bibinfo {author} {\bibfnamefont {M.}~\bibnamefont {de~Souza}},
  \bibinfo {author} {\bibfnamefont {M.~S.}\ \bibnamefont {Figueira}}, \bibinfo
  {author} {\bibfnamefont {I.~A.}\ \bibnamefont {Shelykh}}, \ and\ \bibinfo
  {author} {\bibfnamefont {A.~C.}\ \bibnamefont {Seridonio}},\ }\href {\doibase
  10.1103/PhysRevB.99.155159} {\bibfield  {journal} {\bibinfo  {journal} {Phys.
  Rev. B}\ }\textbf {\bibinfo {volume} {99}},\ \bibinfo {pages} {155159}
  (\bibinfo {year} {2019}{\natexlab{a}})}\BibitemShut {NoStop}%
\bibitem [{\citenamefont {Ricco}\ \emph
  {et~al.}(2019{\natexlab{b}})\citenamefont {Ricco}, \citenamefont {de~Souza},
  \citenamefont {Figueira}, \citenamefont {Shelykh},\ and\ \citenamefont
  {Seridonio}}]{RiccoPRB2019}%
  \BibitemOpen
  \bibfield  {author} {\bibinfo {author} {\bibfnamefont {L.~S.}\ \bibnamefont
  {Ricco}}, \bibinfo {author} {\bibfnamefont {M.}~\bibnamefont {de~Souza}},
  \bibinfo {author} {\bibfnamefont {M.~S.}\ \bibnamefont {Figueira}}, \bibinfo
  {author} {\bibfnamefont {I.~A.}\ \bibnamefont {Shelykh}}, \ and\ \bibinfo
  {author} {\bibfnamefont {A.~C.}\ \bibnamefont {Seridonio}},\ }\href {\doibase
  10.1103/PhysRevB.99.155159} {\bibfield  {journal} {\bibinfo  {journal} {Phys.
  Rev. B}\ }\textbf {\bibinfo {volume} {99}},\ \bibinfo {pages} {155159}
  (\bibinfo {year} {2019}{\natexlab{b}})}\BibitemShut {NoStop}%
\bibitem [{\citenamefont {Bara{\'{n}}ski}\ \emph {et~al.}(2016)\citenamefont
  {Bara{\'{n}}ski}, \citenamefont {Kobia{\l}ka},\ and\ \citenamefont
  {Doma{\'{n}}ski}}]{Baraski2016}%
  \BibitemOpen
  \bibfield  {author} {\bibinfo {author} {\bibfnamefont {J.}~\bibnamefont
  {Bara{\'{n}}ski}}, \bibinfo {author} {\bibfnamefont {A.}~\bibnamefont
  {Kobia{\l}ka}}, \ and\ \bibinfo {author} {\bibfnamefont {T.}~\bibnamefont
  {Doma{\'{n}}ski}},\ }\href {\doibase 10.1088/1361-648x/aa5214} {\bibfield
  {journal} {\bibinfo  {journal} {J. Phys.: Condens. Matter}\ }\textbf
  {\bibinfo {volume} {29}},\ \bibinfo {pages} {075603} (\bibinfo {year}
  {2016})}\BibitemShut {NoStop}%
\bibitem [{\citenamefont {Silva}\ and\ \citenamefont
  {Vernek}(2016)}]{Silva2016}%
  \BibitemOpen
  \bibfield  {author} {\bibinfo {author} {\bibfnamefont {J.~F.}\ \bibnamefont
  {Silva}}\ and\ \bibinfo {author} {\bibfnamefont {E.}~\bibnamefont {Vernek}},\
  }\href {\doibase 10.1088/0953-8984/28/43/435702} {\bibfield  {journal}
  {\bibinfo  {journal} {J. Phys.: Condens. Matter}\ }\textbf {\bibinfo {volume}
  {28}},\ \bibinfo {pages} {435702} (\bibinfo {year} {2016})}\BibitemShut
  {NoStop}%
\bibitem [{\citenamefont {G{\'{o}}rski}\ \emph {et~al.}(2018)\citenamefont
  {G{\'{o}}rski}, \citenamefont {Bara{\'{n}}ski}, \citenamefont {Weymann},\
  and\ \citenamefont {Doma{\'{n}}ski}}]{Gorski2018}%
  \BibitemOpen
  \bibfield  {author} {\bibinfo {author} {\bibfnamefont {G.}~\bibnamefont
  {G{\'{o}}rski}}, \bibinfo {author} {\bibfnamefont {J.}~\bibnamefont
  {Bara{\'{n}}ski}}, \bibinfo {author} {\bibfnamefont {I.}~\bibnamefont
  {Weymann}}, \ and\ \bibinfo {author} {\bibfnamefont {T.}~\bibnamefont
  {Doma{\'{n}}ski}},\ }\href {\doibase 10.1038/s41598-018-33529-1} {\bibfield
  {journal} {\bibinfo  {journal} {Sci. Reports}\ }\textbf {\bibinfo {volume}
  {8}},\ \bibinfo {pages} {15717} (\bibinfo {year} {2018})}\BibitemShut
  {NoStop}%
\bibitem [{\citenamefont {Zienkiewicz}\ \emph {et~al.}(2019)\citenamefont
  {Zienkiewicz}, \citenamefont {Bara{\'n}ski}, \citenamefont {G{\'o}rski},\
  and\ \citenamefont {Doma{\'n}ski}}]{ZienkiewiczarXiv2019}%
  \BibitemOpen
  \bibfield  {author} {\bibinfo {author} {\bibfnamefont {T.}~\bibnamefont
  {Zienkiewicz}}, \bibinfo {author} {\bibfnamefont {J.}~\bibnamefont
  {Bara{\'n}ski}}, \bibinfo {author} {\bibfnamefont {G.}~\bibnamefont
  {G{\'o}rski}}, \ and\ \bibinfo {author} {\bibfnamefont {T.}~\bibnamefont
  {Doma{\'n}ski}},\ }\href@noop {} {\bibfield  {journal} {\bibinfo  {journal}
  {arXiv e-prints}\ ,\ \bibinfo {eid} {arXiv:1908.00349}} (\bibinfo {year}
  {2019})},\ \Eprint {http://arxiv.org/abs/1908.00349} {arXiv:1908.00349
  [cond-mat.mes-hall]} \BibitemShut {NoStop}%
\bibitem [{\citenamefont {Anderson}(1961)}]{Anderson}%
  \BibitemOpen
  \bibfield  {author} {\bibinfo {author} {\bibfnamefont {P.~W.}\ \bibnamefont
  {Anderson}},\ }\href {\doibase 10.1103/PhysRev.124.41} {\bibfield  {journal}
  {\bibinfo  {journal} {Phys. Rev.}\ }\textbf {\bibinfo {volume} {124}},\
  \bibinfo {pages} {41} (\bibinfo {year} {1961})}\BibitemShut {NoStop}%
\bibitem [{\citenamefont {Hoffman}\ \emph {et~al.}(2017)\citenamefont
  {Hoffman}, \citenamefont {Chevallier}, \citenamefont {Loss},\ and\
  \citenamefont {Klinovaja}}]{Hoffman2017}%
  \BibitemOpen
  \bibfield  {author} {\bibinfo {author} {\bibfnamefont {S.}~\bibnamefont
  {Hoffman}}, \bibinfo {author} {\bibfnamefont {D.}~\bibnamefont {Chevallier}},
  \bibinfo {author} {\bibfnamefont {D.}~\bibnamefont {Loss}}, \ and\ \bibinfo
  {author} {\bibfnamefont {J.}~\bibnamefont {Klinovaja}},\ }\href {\doibase
  10.1103/PhysRevB.96.045440} {\bibfield  {journal} {\bibinfo  {journal} {Phys.
  Rev. B}\ }\textbf {\bibinfo {volume} {96}},\ \bibinfo {pages} {045440}
  (\bibinfo {year} {2017})}\BibitemShut {NoStop}%
\bibitem [{\citenamefont {Ricco}\ \emph
  {et~al.}(2018{\natexlab{a}})\citenamefont {Ricco}, \citenamefont {Dessotti},
  \citenamefont {Shelykh}, \citenamefont {Figueira},\ and\ \citenamefont
  {Seridonio}}]{RiccoSciReports}%
  \BibitemOpen
  \bibfield  {author} {\bibinfo {author} {\bibfnamefont {L.~S.}\ \bibnamefont
  {Ricco}}, \bibinfo {author} {\bibfnamefont {F.~A.}\ \bibnamefont {Dessotti}},
  \bibinfo {author} {\bibfnamefont {I.~A.}\ \bibnamefont {Shelykh}}, \bibinfo
  {author} {\bibfnamefont {M.~S.}\ \bibnamefont {Figueira}}, \ and\ \bibinfo
  {author} {\bibfnamefont {A.~C.}\ \bibnamefont {Seridonio}},\ }\href {\doibase
  10.1038/s41598-018-21180-9} {\bibfield  {journal} {\bibinfo  {journal} {Sci.
  Reports}\ }\textbf {\bibinfo {volume} {8}},\ \bibinfo {pages} {2790}
  (\bibinfo {year} {2018}{\natexlab{a}})}\BibitemShut {NoStop}%
\bibitem [{\citenamefont {Liu}\ and\ \citenamefont
  {Baranger}(2011)}]{Baranger2011}%
  \BibitemOpen
  \bibfield  {author} {\bibinfo {author} {\bibfnamefont {D.~E.}\ \bibnamefont
  {Liu}}\ and\ \bibinfo {author} {\bibfnamefont {H.~U.}\ \bibnamefont
  {Baranger}},\ }\href {\doibase 10.1103/PhysRevB.84.201308} {\bibfield
  {journal} {\bibinfo  {journal} {Phys. Rev. B}\ }\textbf {\bibinfo {volume}
  {84}},\ \bibinfo {pages} {201308} (\bibinfo {year} {2011})}\BibitemShut
  {NoStop}%
\bibitem [{\citenamefont {Ricco}\ \emph
  {et~al.}(2018{\natexlab{b}})\citenamefont {Ricco}, \citenamefont {Campo},
  \citenamefont {Shelykh},\ and\ \citenamefont
  {Seridonio}}]{RiccoOscillations2018}%
  \BibitemOpen
  \bibfield  {author} {\bibinfo {author} {\bibfnamefont {L.~S.}\ \bibnamefont
  {Ricco}}, \bibinfo {author} {\bibfnamefont {V.~L.}\ \bibnamefont {Campo}},
  \bibinfo {author} {\bibfnamefont {I.~A.}\ \bibnamefont {Shelykh}}, \ and\
  \bibinfo {author} {\bibfnamefont {A.~C.}\ \bibnamefont {Seridonio}},\ }\href
  {\doibase 10.1103/PhysRevB.98.075142} {\bibfield  {journal} {\bibinfo
  {journal} {Phys. Rev. B}\ }\textbf {\bibinfo {volume} {98}},\ \bibinfo
  {pages} {075142} (\bibinfo {year} {2018}{\natexlab{b}})}\BibitemShut
  {NoStop}%
\bibitem [{\citenamefont {G{\'{o}}rski}\ and\ \citenamefont
  {Kucab}(2019)}]{Gorski2019}%
  \BibitemOpen
  \bibfield  {author} {\bibinfo {author} {\bibfnamefont {G.}~\bibnamefont
  {G{\'{o}}rski}}\ and\ \bibinfo {author} {\bibfnamefont {K.}~\bibnamefont
  {Kucab}},\ }\href {\doibase 10.1002/pssb.201800492} {\bibfield  {journal}
  {\bibinfo  {journal} {Phys. Status Solidi B}\ }\textbf {\bibinfo {volume}
  {256}},\ \bibinfo {pages} {1800492} (\bibinfo {year} {2019})}\BibitemShut
  {NoStop}%
\bibitem [{\citenamefont {Tanaka}\ \emph {et~al.}(2007)\citenamefont {Tanaka},
  \citenamefont {Kawakami},\ and\ \citenamefont {Oguri}}]{Tanaka2007}%
  \BibitemOpen
  \bibfield  {author} {\bibinfo {author} {\bibfnamefont {Y.}~\bibnamefont
  {Tanaka}}, \bibinfo {author} {\bibfnamefont {N.}~\bibnamefont {Kawakami}}, \
  and\ \bibinfo {author} {\bibfnamefont {A.}~\bibnamefont {Oguri}},\ }\href
  {\doibase 10.1143/JPSJ.76.074701} {\bibfield  {journal} {\bibinfo  {journal}
  {J. Phys. Soc. Jpn.}\ }\textbf {\bibinfo {volume} {76}},\ \bibinfo {pages}
  {074701} (\bibinfo {year} {2007})}\BibitemShut {NoStop}%
\bibitem [{\citenamefont {Bara{\'{n}}ski}\ and\ \citenamefont
  {Doma{\'{n}}ski}(2013)}]{Baraski2013}%
  \BibitemOpen
  \bibfield  {author} {\bibinfo {author} {\bibfnamefont {J.}~\bibnamefont
  {Bara{\'{n}}ski}}\ and\ \bibinfo {author} {\bibfnamefont {T.}~\bibnamefont
  {Doma{\'{n}}ski}},\ }\href {\doibase 10.1088/0953-8984/25/43/435305}
  {\bibfield  {journal} {\bibinfo  {journal} {J. Phys.: Condens. Matter}\
  }\textbf {\bibinfo {volume} {25}},\ \bibinfo {pages} {435305} (\bibinfo
  {year} {2013})}\BibitemShut {NoStop}%
\bibitem [{\citenamefont {Oguri}\ \emph {et~al.}(2004)\citenamefont {Oguri},
  \citenamefont {Tanaka},\ and\ \citenamefont {C.~Hewson}}]{Oguri2004}%
  \BibitemOpen
  \bibfield  {author} {\bibinfo {author} {\bibfnamefont {A.}~\bibnamefont
  {Oguri}}, \bibinfo {author} {\bibfnamefont {Y.}~\bibnamefont {Tanaka}}, \
  and\ \bibinfo {author} {\bibfnamefont {A.}~\bibnamefont {C.~Hewson}},\ }\href
  {\doibase 10.1143/JPSJ.73.2494} {\bibfield  {journal} {\bibinfo  {journal}
  {J. Phys. Soc. Jpn.}\ }\textbf {\bibinfo {volume} {73}},\ \bibinfo {pages}
  {2494} (\bibinfo {year} {2004})}\BibitemShut {NoStop}%
\bibitem [{\citenamefont {Bauer}\ \emph {et~al.}(2007)\citenamefont {Bauer},
  \citenamefont {Oguri},\ and\ \citenamefont {Hewson}}]{Bauer2007}%
  \BibitemOpen
  \bibfield  {author} {\bibinfo {author} {\bibfnamefont {J.}~\bibnamefont
  {Bauer}}, \bibinfo {author} {\bibfnamefont {A.}~\bibnamefont {Oguri}}, \ and\
  \bibinfo {author} {\bibfnamefont {A.~C.}\ \bibnamefont {Hewson}},\ }\href
  {\doibase 10.1088/0953-8984/19/48/486211} {\bibfield  {journal} {\bibinfo
  {journal} {J. Phys.: Condens. Matter}\ }\textbf {\bibinfo {volume} {19}},\
  \bibinfo {pages} {486211} (\bibinfo {year} {2007})}\BibitemShut {NoStop}%
\bibitem [{\citenamefont {Martín-Rodero}\ and\ \citenamefont
  {Yeyati}(2011)}]{Rodero2011}%
  \BibitemOpen
  \bibfield  {author} {\bibinfo {author} {\bibfnamefont {A.}~\bibnamefont
  {Martín-Rodero}}\ and\ \bibinfo {author} {\bibfnamefont {A.~L.}\
  \bibnamefont {Yeyati}},\ }\href {\doibase 10.1080/00018732.2011.624266}
  {\bibfield  {journal} {\bibinfo  {journal} {Advances in Physics}\ }\textbf
  {\bibinfo {volume} {60}},\ \bibinfo {pages} {899} (\bibinfo {year}
  {2011})}\BibitemShut {NoStop}%
\bibitem [{\citenamefont {Ma\ifmmode~\acute{s}\else \'{s}\fi{}ka}\ \emph
  {et~al.}(2017)\citenamefont {Ma\ifmmode~\acute{s}\else \'{s}\fi{}ka},
  \citenamefont {Gorczyca-Goraj}, \citenamefont {Tworzyd\l{}o},\ and\
  \citenamefont {Doma\ifmmode~\acute{n}\else \'{n}\fi{}ski}}]{MaskaPRB2017}%
  \BibitemOpen
  \bibfield  {author} {\bibinfo {author} {\bibfnamefont {M.~M.}\ \bibnamefont
  {Ma\ifmmode~\acute{s}\else \'{s}\fi{}ka}}, \bibinfo {author} {\bibfnamefont
  {A.}~\bibnamefont {Gorczyca-Goraj}}, \bibinfo {author} {\bibfnamefont
  {J.}~\bibnamefont {Tworzyd\l{}o}}, \ and\ \bibinfo {author} {\bibfnamefont
  {T.}~\bibnamefont {Doma\ifmmode~\acute{n}\else \'{n}\fi{}ski}},\ }\href
  {\doibase 10.1103/PhysRevB.95.045429} {\bibfield  {journal} {\bibinfo
  {journal} {Phys. Rev. B}\ }\textbf {\bibinfo {volume} {95}},\ \bibinfo
  {pages} {045429} (\bibinfo {year} {2017})}\BibitemShut {NoStop}%
\bibitem [{Note1()}]{Note1}%
  \BibitemOpen
  \bibinfo {note} {Away from Kondo regime~\cite {Baraski2016,LeePRB,Zitko2015},
  the effects of Coulomb blockade in the energy spectrum of the QD coupled to
  both S and N leads are well-described within the following self-consistent
  Hartree-Fock approximation~\cite {Bruus,LeeNanotech,Prada2017,Rodero2012}:
  $Un_{d\protect \downarrow }n_{d\protect \uparrow }\approx U(\protect \langle
  n_{d\protect \downarrow }\protect \rangle n_{d\protect \uparrow
  }+n_{d\protect \downarrow }\protect \langle n_{d\protect \uparrow }\protect
  \rangle -\protect \langle d_{\protect \downarrow }d_{\protect \uparrow
  }\protect \rangle d_{\protect \uparrow }^{\dagger }d_{\protect \downarrow
  }^{\dagger } -\protect \langle d_{\protect \uparrow }^{\dagger }d_{\protect
  \downarrow }^{\dagger }\protect \rangle d_{\protect \downarrow }d_{\protect
  \uparrow }) + \protect \text {const.}$, where $\protect \langle n{}_{d\sigma
  }\protect \rangle =(-\protect \frac {1}{\pi })\DOTSI \intop \ilimits@
  _{-\infty }^{0}d\omega \protect \text {Im}[\protect \langle \protect \langle
  d_{\sigma } d_{\sigma }^{\dagger }\protect \rangle \protect \rangle ]$ and
  $\protect \langle d_{\protect \bar {\sigma }}^{\dagger }d_{\sigma }^{\dagger
  }\protect \rangle =(-\protect \frac {1}{\pi })\DOTSI \intop \ilimits@
  _{-\infty }^{0}d\omega \protect \text {Im}[\protect \langle \protect \langle
  d_{\protect \bar {\sigma }}^{\dagger };d_{\sigma }^{\dagger }\protect \rangle
  \protect \rangle ]$ are the average occupation and \protect \textit {s}-wave
  pairing in the QD, respectively. Both quantities were computed with
  self-consistent numerical calculations.}\BibitemShut {Stop}%
\bibitem [{\citenamefont {Andreev}(1964)}]{Andreev1964}%
  \BibitemOpen
  \bibfield  {author} {\bibinfo {author} {\bibfnamefont {A.~F.}\ \bibnamefont
  {Andreev}},\ }\href
  {http://www.jetp.ac.ru/cgi-bin/e/index/e/19/5/p1228?a=list} {\bibfield
  {journal} {\bibinfo  {journal} {J. Exp. Teor. Phys}\ }\textbf {\bibinfo
  {volume} {19}},\ \bibinfo {pages} {1228} (\bibinfo {year}
  {1964})}\BibitemShut {NoStop}%
\bibitem [{\citenamefont {Krawiec}\ and\ \citenamefont
  {Wysoki{\'{n}}ski}(2003)}]{Krawiec2003}%
  \BibitemOpen
  \bibfield  {author} {\bibinfo {author} {\bibfnamefont {M.}~\bibnamefont
  {Krawiec}}\ and\ \bibinfo {author} {\bibfnamefont {K.~I.}\ \bibnamefont
  {Wysoki{\'{n}}ski}},\ }\href {\doibase 10.1088/0953-2048/17/1/018} {\bibfield
   {journal} {\bibinfo  {journal} {Supercond. Sci. Technol.}\ }\textbf
  {\bibinfo {volume} {17}},\ \bibinfo {pages} {103} (\bibinfo {year}
  {2003})}\BibitemShut {NoStop}%
\bibitem [{\citenamefont {Haug}\ and\ \citenamefont {Jauho}(2008)}]{Jauho}%
  \BibitemOpen
  \bibfield  {author} {\bibinfo {author} {\bibfnamefont {H.}~\bibnamefont
  {Haug}}\ and\ \bibinfo {author} {\bibfnamefont {A.}~\bibnamefont {Jauho}},\
  }\href@noop {} {\emph {\bibinfo {title} {Quantum Kinetics in Transport and
  Optics of Semiconductors}}},\ Springer Series in Solid-State Sciences\
  (\bibinfo  {publisher} {Springer Berlin Heidelberg},\ \bibinfo {year}
  {2008})\BibitemShut {NoStop}%
\bibitem [{\citenamefont {Bruus}\ and\ \citenamefont
  {Flensberg}(2004)}]{Bruus}%
  \BibitemOpen
  \bibfield  {author} {\bibinfo {author} {\bibfnamefont {H.}~\bibnamefont
  {Bruus}}\ and\ \bibinfo {author} {\bibfnamefont {K.}~\bibnamefont
  {Flensberg}},\ }\href@noop {} {\emph {\bibinfo {title} {Many-Body Quantum
  Theory in Condensed Matter Physics: An Introduction}}},\ Oxford Graduate
  Texts\ (\bibinfo  {publisher} {Oxford University Press},\ \bibinfo {year}
  {2004})\BibitemShut {NoStop}%
\bibitem [{Note2()}]{Note2}%
  \BibitemOpen
  \bibinfo {note} {It is worth mentioning that Eq.~(\ref {eq:MatrixGsigma}) has
  similar shape of those found by Zienkiewicz \protect \textit {et al.}~\cite
  {ZienkiewiczarXiv2019}. Such a matrix type also was computed by G\'{o}rski
  and Kucab~\cite {Gorski2019} without the S reservoir and by Ramos-Andrade
  \protect \textit {et al.}~\cite {RamosAndrade2019} for a QD between N leads
  and side coupled to two TSC nanowires.}\BibitemShut {Stop}%
\bibitem [{\citenamefont {Vernek}\ \emph {et~al.}(2014)\citenamefont {Vernek},
  \citenamefont {Penteado}, \citenamefont {Seridonio},\ and\ \citenamefont
  {Egues}}]{VernekPRB2014}%
  \BibitemOpen
  \bibfield  {author} {\bibinfo {author} {\bibfnamefont {E.}~\bibnamefont
  {Vernek}}, \bibinfo {author} {\bibfnamefont {P.~H.}\ \bibnamefont
  {Penteado}}, \bibinfo {author} {\bibfnamefont {A.~C.}\ \bibnamefont
  {Seridonio}}, \ and\ \bibinfo {author} {\bibfnamefont {J.~C.}\ \bibnamefont
  {Egues}},\ }\href {\doibase 10.1103/PhysRevB.89.165314} {\bibfield  {journal}
  {\bibinfo  {journal} {Phys. Rev. B}\ }\textbf {\bibinfo {volume} {89}},\
  \bibinfo {pages} {165314} (\bibinfo {year} {2014})}\BibitemShut {NoStop}%
\bibitem [{Note3()}]{Note3}%
  \BibitemOpen
  \bibinfo {note} {Experimentally, these quantities can be continuously tuned
  by setting up a dual-gate geometry, with the insertion of a global back-gate
  in the setup, as performed by E. J. H. Lee \protect \textit {et al.}~\cite
  {LeePRB}.}\BibitemShut {Stop}%
\bibitem [{\citenamefont {Lee}\ \emph {et~al.}(2017)\citenamefont {Lee},
  \citenamefont {Jiang}, \citenamefont {\ifmmode~\check{Z}\else
  \v{Z}\fi{}itko}, \citenamefont {Aguado}, \citenamefont {Lieber},\ and\
  \citenamefont {De~Franceschi}}]{LeePRB}%
  \BibitemOpen
  \bibfield  {author} {\bibinfo {author} {\bibfnamefont {E.~J.~H.}\
  \bibnamefont {Lee}}, \bibinfo {author} {\bibfnamefont {X.}~\bibnamefont
  {Jiang}}, \bibinfo {author} {\bibfnamefont {R.}~\bibnamefont
  {\ifmmode~\check{Z}\else \v{Z}\fi{}itko}}, \bibinfo {author} {\bibfnamefont
  {R.}~\bibnamefont {Aguado}}, \bibinfo {author} {\bibfnamefont {C.~M.}\
  \bibnamefont {Lieber}}, \ and\ \bibinfo {author} {\bibfnamefont
  {S.}~\bibnamefont {De~Franceschi}},\ }\href {\doibase
  10.1103/PhysRevB.95.180502} {\bibfield  {journal} {\bibinfo  {journal} {Phys.
  Rev. B}\ }\textbf {\bibinfo {volume} {95}},\ \bibinfo {pages} {180502}
  (\bibinfo {year} {2017})}\BibitemShut {NoStop}%
\bibitem [{\citenamefont {\ifmmode~\check{Z}\else \v{Z}\fi{}itko}\ \emph
  {et~al.}(2015)\citenamefont {\ifmmode~\check{Z}\else \v{Z}\fi{}itko},
  \citenamefont {Lim}, \citenamefont {L\'opez},\ and\ \citenamefont
  {Aguado}}]{Zitko2015}%
  \BibitemOpen
  \bibfield  {author} {\bibinfo {author} {\bibfnamefont {R.}~\bibnamefont
  {\ifmmode~\check{Z}\else \v{Z}\fi{}itko}}, \bibinfo {author} {\bibfnamefont
  {J.~S.}\ \bibnamefont {Lim}}, \bibinfo {author} {\bibfnamefont
  {R.}~\bibnamefont {L\'opez}}, \ and\ \bibinfo {author} {\bibfnamefont
  {R.}~\bibnamefont {Aguado}},\ }\href {\doibase 10.1103/PhysRevB.91.045441}
  {\bibfield  {journal} {\bibinfo  {journal} {Phys. Rev. B}\ }\textbf {\bibinfo
  {volume} {91}},\ \bibinfo {pages} {045441} (\bibinfo {year}
  {2015})}\BibitemShut {NoStop}%
\bibitem [{\citenamefont {Lee}\ \emph {et~al.}(2013)\citenamefont {Lee},
  \citenamefont {Jiang}, \citenamefont {Houzet}, \citenamefont {Aguado},
  \citenamefont {Lieber},\ and\ \citenamefont {De~Franceschi}}]{LeeNanotech}%
  \BibitemOpen
  \bibfield  {author} {\bibinfo {author} {\bibfnamefont {E.~J.~H.}\
  \bibnamefont {Lee}}, \bibinfo {author} {\bibfnamefont {X.}~\bibnamefont
  {Jiang}}, \bibinfo {author} {\bibfnamefont {M.}~\bibnamefont {Houzet}},
  \bibinfo {author} {\bibfnamefont {R.}~\bibnamefont {Aguado}}, \bibinfo
  {author} {\bibfnamefont {C.~M.}\ \bibnamefont {Lieber}}, \ and\ \bibinfo
  {author} {\bibfnamefont {S.}~\bibnamefont {De~Franceschi}},\ }\href {\doibase
  10.1038/nnano.2013.267} {\bibfield  {journal} {\bibinfo  {journal} {Nature
  Nanotech.}\ }\textbf {\bibinfo {volume} {9}},\ \bibinfo {pages} {79}
  (\bibinfo {year} {2013})}\BibitemShut {NoStop}%
\bibitem [{\citenamefont {Mart{\'{\i}}n-Rodero}\ and\ \citenamefont
  {Yeyati}(2012)}]{Rodero2012}%
  \BibitemOpen
  \bibfield  {author} {\bibinfo {author} {\bibfnamefont {A.}~\bibnamefont
  {Mart{\'{\i}}n-Rodero}}\ and\ \bibinfo {author} {\bibfnamefont {A.~L.}\
  \bibnamefont {Yeyati}},\ }\href {\doibase 10.1088/0953-8984/24/38/385303}
  {\bibfield  {journal} {\bibinfo  {journal} {J. Phys.: Condens. Matter}\
  }\textbf {\bibinfo {volume} {24}},\ \bibinfo {pages} {385303} (\bibinfo
  {year} {2012})}\BibitemShut {NoStop}%
\bibitem [{\citenamefont {Ramos-Andrade}\ \emph {et~al.}(2019)\citenamefont
  {Ramos-Andrade}, \citenamefont {Zambrano},\ and\ \citenamefont
  {Orellana}}]{RamosAndrade2019}%
  \BibitemOpen
  \bibfield  {author} {\bibinfo {author} {\bibfnamefont {J.~P.}\ \bibnamefont
  {Ramos-Andrade}}, \bibinfo {author} {\bibfnamefont {D.}~\bibnamefont
  {Zambrano}}, \ and\ \bibinfo {author} {\bibfnamefont {P.~A.}\ \bibnamefont
  {Orellana}},\ }\href {\doibase 10.1002/andp.201800498} {\bibfield  {journal}
  {\bibinfo  {journal} {Annalen der Physik}\ }\textbf {\bibinfo {volume}
  {531}},\ \bibinfo {pages} {1800498} (\bibinfo {year} {2019})}\BibitemShut
  {NoStop}%
\end{thebibliography}%

\end{document}